\documentclass{aa}  

\usepackage{graphicx}
\usepackage{amsmath,amssymb}
\usepackage{longtable,lscape}
\usepackage{txfonts}
\begin{document}

   \title{SABOCA 350-$\mu$m and LABOCA 870-$\mu$m dust continuum imaging of 
IRAS 05399-0121: mapping the dust properties of a pre- and protostellar core 
system\thanks{This publication is based on data acquired with the Atacama 
Pathfinder EXperiment (APEX) under programmes 
079.F-9313(A) and 090.F-9307(A). APEX is a collaboration between the 
Max-Planck-Institut f\"{u}r Radioastronomie, the European Southern 
Observatory, and the Onsala Space Observatory.}}

%   \subtitle{}

   \author{O. Miettinen\inst{1} and S.~S.~R. Offner\inst{2}\thanks{Hubble Postdoctoral Fellow.}}

   \institute{Department of Physics, University of Helsinki, P.O. Box 64, FI-00014 Helsinki, Finland\\ \email{oskari.miettinen@helsinki.fi} \and Department of 
Astronomy, Yale University, New Haven, CT 06511, USA \\ \email{stella.offner@yale.edu}}

   \date{Received ; accepted}

% \abstract{}{}{}{}{} 
% 5 {} token are mandatory
\authorrunning{Miettinen \& Offner}
\titlerunning{SABOCA and LABOCA imaging of IRAS 05399-0121}

  \abstract
  % context heading (optional)
  % {} leave it empty if necessary  
   {Thermal emission from dust provides a valuable tool for determining 
important physical properties of the dense structures within mole\-cular 
clouds.}
  % aims heading (mandatory)
   {We attempt to map the distributions of dust temperature and H$_2$ column 
density of IRAS 05399-0121/SMM 1, which is a dense double-core system in Orion 
B9. We also search for substructures within the cores through high-resolution 
submillimetre imaging.}
  % methods heading (mandatory)
   {The source was mapped with APEX/SABOCA at 350 $\mu$m. We combined these 
data with our previous LABOCA 870-$\mu$m data. The spatial resolution of the 
new SABOCA image, $\sim3\,400$ AU, is about 2.6 times better than provided by 
LABOCA, and is therefore well-suited to our purposes. We also make use of the 
\textit{Spitzer} infrared observations to characterise the star-formation 
activity in the source.}
  % results heading (mandatory)
   {The filamentary source remains a double-core system on the 3\,400 AU scale 
probed here, and the projected separation between IRAS 05399 and SMM 1 is 0.14 
pc. The temperature map reveals warm spots towards IRAS 05399 and the 
southeastern tip of the source. Both IRAS 05399 and SMM 1 stand out as peaks 
in the column density map. A simple analysis suggests that the density profile 
has the form $\sim r^{-(2.3_{-0.9}^{+2.2})}$, as determined at the 
position of SMM 1. The broadband spectral energy distribution of IRAS 05399 
suggests that it is near the Stage 0/I borderline. A visual inspection of the 
\textit{Spitzer}/IRAC images provides hints of a quadrupolar-like jet 
morphology around IRAS 05399, supporting the possibility that it is a binary 
system.}
  % conclusions heading (optional), leave it empty if necessary 
   {The source splitting into two subcores along the long axis can be 
explained by cylindrical Jeans-type fragmentation, but the steepness of the 
density profile is shallower than what is expected for an isothermal 
cylinder. The difference between the evolutionary stages of IRAS 05399 
(protostellar) and SMM 1 (starless) suggests that the former has experienced a 
phase of rapid mass accretion, supported by the very long outflow it drives. 
%This could be a manifestation of triggered star formation in Orion B9. 
The protostellar jet from IRAS 05399 might have influenced the nearby 
core SMM 1. In particular, the temperature map features are likely to be 
imprints of protostellar or shock heating, while external heating could be 
provided by the nearby high-mass star-forming region NGC 2024.}


   \maketitle
%
%________________________________________________________________

\section{Introduction}

Understanding of Galactic star formation is 
ultimately linked to the understanding of the physical properties of dense 
cores in molecular clouds, because they appear to be the direct proge\-nitors 
of new stars. It is well established that the formation process 
of low-mass ($\sim0.1-2$ M$_{\sun}$) solar-type stars start in 
gra\-vitationally bound starless cores or so-called prestellar cores 
(\cite{wardthompson1994}). Because these sources re\-present the physical 
conditions of dense gas and dust best before the influence of star formation, 
they are especially useful targets for studies of the initial stages of the 
star-formation process. At some point, a prestellar core collapses due to 
its self-gravity, and, after a short ($\sim10^2-10^3$ yr) transitional 
first-hydrostatic core stage (\cite{larson1969}; \cite{masunaga1998}; 
\cite{bate2011}), a central protostar forms in its centre 
(the so-called second hydrostatic core; e.g., \cite{masunaga2000}). 
The emission of these young accreting sources is heavily 
reprocessed by the dense embedding gas and dust, and consequently they appear 
as cold, sub-mm sources also known as Class 0 objects (\cite{andre1993}, 2000).
%At this stage, most of the system's mass is 
%still in the dense envelope surrounding the protostar ($M_{\rm env}\gg M_{\star}$, where $M_{\star}$ is the mass of the central protostar). 

Based on statistical estimates, the lifetime of the 
Class 0 stage is $\sim1\times10^5$ yr (\cite{evans2009}; \cite{enoch2009}). 
However, the accretion rate and duration can be highly dependent on 
the initial and/or environmental conditions (e.g., \cite{vorobyov2010}; 
\cite{offner2012b}). Over time, the protostellar envelope dissipates 
through mass accretion onto the forming star and ejection of circumstellar 
material through outflows (e.g., \cite{arce2006}). The source emission 
shifts to shorter wavelengths, peaking in the infrared, and the object becomes 
a Class I source. The transition from Class 0 to Class I is thought to occur 
when the envelope mass is approximately equal to the mass of the central 
protostar. However, protostellar masses cannot be directly measured, and in 
practice source classification is very sensitive to the inclination of the 
protostellar disk and outflow (e.g., \cite{whitney2003}; 
\cite{robitaille2007}; \cite{offner2012b}). After the envelope is dispersed, 
further accretion is minimal, the source is optically revealed, and it enters 
the pre-main sequence (e.g., \cite{wuchterl2003}).
%In any case, as time passes, the protostellar 
%envelope dissipates through mass accretion onto the forming star and ejection 
%of circumstellar material through outflows (e.g., \cite{arce2006}). 
%When the envelope mass is about 
%equal to the mass of the central protostar, $M_{\rm env}\simeq M_{\star}$, 
%the transition phase between Class 0 and I is said to take place (commonly 
%designated as Class 0/I). The protostars in the Class I stage have already 
%accumulated the majority of their final stellar mass ($M_{\star}>M_{\rm env}$; 
%e.g., \cite{wilking1989}). 
%The Class I stage represents the last protostellar 
%phase of low-mass star formation, and after that the source enters the 
%pre-main sequence (e.g., \cite{wuchterl2003}).

The target source of the present study is a dense core system in the Orion 
B9 star-forming region that consists of a candidate prestellar core and a 
Class 0/I protostar, namely the \textit{IRAS} (\textit{Infrared Astronomical 
Satellite}; \cite{neugebauer1984}) source 05399-0121 (hereafter, I05399). 
This source has been the target of se\-veral previous studies, and what 
follows is an overview of these studies in chronological order. 

Wouterloot \& Walmsley (1986) searched for 22-GHz 
H$_2$O maser emission from I05399 but it was not detected. The source was 
also included in the NH$_3$ survey by Wouterloot et al. (1988), where the 
$(1,\,1)$ line was detected at 9.25 km~s$^{-1}$ but only an upper limit to the 
gas kinetic temperature of $T_{\rm kin}<11.1$ K could be derived. The same 
authors later included I05399 as part of their $^{12}$CO/$^{13}$CO survey 
of selected \textit{IRAS} sources (\cite{wouterloot1989}). As can be seen in 
their Fig.~1, the $^{12}$CO$(1-0)$ line shows a blue asymmetric profile with 
the blue peak being 4 K stronger than the red peak. Moreover, the 
$^{13}$CO$(1-0)$ line peaks near the central dip of the double-peaked 
$^{12}$CO line. This could be the signature of the infalling gas 
motions (e.g., \cite{zhou1992}; \cite{myers1996}). However, it is also 
possible that there are just distinct velocity components along the line of 
sight; for example, there is additional CO isotopologue emission at 
$\sim2.5-3$ km~s$^{-1}$ (\cite{wouterloot1989}; their Fig.~1). In 
addition, the $^{12}$CO line detected by these authors shows 
non-Gaussian, high-velocity wing emission of low intensity indicative of 
outflowing gas. 

Lada et al. (1991) performed a CS$(2-1)$ survey of the Orion B molecular 
cloud. The CS clump they found to be associated with I05399 (No.~30) is 
commonly called LBS 30 according to authors' last names. Within the $5\sigma$ 
emission level, the effective radius and mass of LBS 30 were determined to be 
0.21 pc and 85.6 M$_{\sun}$ (they adopted a distance of 400 pc\footnote{In 
the present paper, we adopt a distance of 450 pc to the Orion giant molecular 
cloud (\cite{genzel1989}). The actual distance may be somewhat smaller as, 
for example, Menten et al. (2007) determined a trigonometric parallax 
distance of $414\pm7$ pc to the Orion Nebula. In this introduction section, 
however, we quote the distance values (and distance-dependent physical 
properties) used by different authors.}). Further NH$_3$ stu\-dies of I05399 
were performed by Harju et al. (1991, 1993). The NH$_3 (1,\,1)$ map from 
Harju et al. (1993; their Appendix) shows an elongated structure 
where an additional ammonia condensation can be seen southeast of I05399. 
The average linewidth and $T_{\rm kin}$ they derived for the mapped source are 
$0.6\pm0.2$ km~s$^{-1}$ and $13.8\pm3.6$ K, respectively, and the mass, as 
estimated from NH$_3$, is 14 M$_{\sun}$ (they adopted a distance of 500 pc). 
Caselli \& Myers (1995) mapped the Orion B9 region in the $J=1-0$ line of 
$^{13}$CO and C$^{18}$O, and I05399 was found 
to lie within the C$^{18}$O clump in the centre of the map (their Fig.~4). 
I05399 was also included in the large 22-GHz H$_2$O maser survey by Codella 
et al. (2002) but, similarly to Wouterloot \& Walmsley (1986), no emission was 
detected. To our knowledge, the first millimetre dust continuum emission map 
of LBS 30 was made by Launhardt et al. (1996). They used both the SEST 15-m 
and IRAM 30-m telescopes to image the 1.3-mm dust emission of LBS 30, and 
found a dust core associated with I05399 and a subcore southeast of it, 
resembling the morphology seen in the NH$_3$ map by Harju et al. (1993). 
Assuming a distance of 400 pc and dust temperature of $T_{\rm dust}=30$ K, 
Launhardt et al. (1996) estimated the total mass of the source to be 3.7 
M$_{\sun}$. LBS 30 was included in the multitransition CS study of Lada et al. 
(1997). Their large velocity gradient modelling of the lines yielded a density 
of $2\times10^5$ cm$^{-3}$ within the region where CS$(5-4)$ line was detected, 
i.e., within the effective radius 0.03 pc at $d=400$ pc. Bergin et al. (1999) 
mapped the source in the $J=1-0$ line of the molecular ions N$_2$H$^+$, 
H$^{13}$CO$^+$, and DCO$^+$, and also in C$^{18}$O$(1-0)$ and CS$(2-1)$ 
(see their Fig.~1). Interestingly, the line emission of all the species 
peaks towards the southeastern core, rather than towards I05399. 
Particularly their N$_2$H$^+$ map appears very similar to that of NH$_3$ 
(\cite{harju1993}), which might not be surprising given that both species 
are excellent tracers of the dense interstellar gas (e.g., \cite{bergin1997}). 
By utilising the molecular ion data, Bergin et al. (1999) constrained the 
fractional ionisation in I05399 to lie in the range 
$9.3\times10^{-8}<x({\rm e})<1.8\times10^{-7}$, with the best-fit value being 
$\sim1.3\times10^{-7}$. I05399 also appeared in the 
C$^{18}$O/H$^{13}$CO$^+$ survey of Aoyama et al. (2001). They designated the 
source as the H$^{13}$CO$^+$ clump No.~7, and derived the effective clump 
radius and mass of 0.23 pc and 18 M$_{\sun}$ (assuming $d=400$ pc). 

One of the most intriguing features of I05399 is the highly 
collimated HH 92 jet emanating from it, and the associated parsec-scale 
bipolar outflow discovered by Bally et al. (2002). This giant 
outflow, consisting of the Herbig-Haro objects HH 90, 91, 92, 93, 597, and 
598, has a projected length of about $34\arcmin$ (4.1 pc at 415 pc).
The HH 92 jet was found to exhibit small-scale wiggles, and HH 598 at the 
northwestern end of the outflow was found to deviate from the current
orientation of the HH 92 jet by $3\degr$ in projection. One explanation for 
the observed features is that I05399 is composed of a binary protostar whose 
orbital period modulates the outflow's mass-loss rate and velocity. 
Connelley et al. (2007) observed 2-$\mu$m emission features extending 
$1\arcmin$ to the northwest of I05399. 
These are likely to be the result of shocked H$_2$ emission from 
the HH 92 jet. I05399 was included in the 22-GHz H$_2$O maser survey of 
\textit{IRAS} sources by Sunada et al. (2007) but, similarly to the above 
mentioned studies, no maser emission was detected. This is consistent with 
the fact that among low-mass young stellar objects (YSOs), 22-GHz H$_2$O 
masers, which are collisionally pumped in shocked gas, are found to be 
associated with predominantly Class 0 protostars and in excited shocks of 
protostellar jets in the close vicinity of the central star (e.g., 
\cite{furuya2003}). I05399 may have already passed through the H$_2$O maser 
phase, but also viewing geometry and variability could play a role.

I05399 was designated as the source J054227.9-012003 in the SCUBA Legacy 
Catalogues by Di Francesco et al. (2008). The 450- and 850-$\mu$m flux 
densities they reported are 26.06 and 3.62 Jy, respectively. 
Ikeda et al. (2009) performed an H$^{13}$CO$^{+}(1-0)$ survey of dense cores 
in Orion B, and their core No.~56 is associated with I05399. The radius 
and mass they derived for this H$^{13}$CO$^{+}$ core are $52\farcs6$ (0.12 pc) 
and 6.8 M$_{\sun}$ (at $d=470$ pc as they assumed). The virial mass they 
derived, 5.8 M$_{\sun}$, suggests that the H$^{13}$CO$^{+}$ core is near 
virial equilibrium. More recently, Miettinen et al. (2009; hereafter, Paper 
I) mapped the whole Orion B9 region in the 870-$\mu$m dust continuum emission 
with the LABOCA (Large APEX BOlometer CAmera) bolometer on APEX. With the peak 
intensity of 0.81 Jy~beam$^{-1}$, I05399 was found to be the second strongest 
870-$\mu$m source in the region. Similarly to earlier studies, 
a dense starless core southeast of I05399 was detected. We suggested 
that it is a prestellar core and called it SMM 1. We also constructed the 
spectral energy distribution (SED) of I05399 by combining data from 
\textit{IRAS}, \textit{Spitzer} (24 and 70 $\mu$m), and LABOCA. 
The dust temperature and mass of the cold envelope, and the bolometric 
luminosity were found to be $18.5\pm0.1$ K, 
$2.8\pm0.3$ M$_{\sun}$, and $21\pm1.2$ L$_{\sun}$, respectively ($d=450$ pc). 
Based on its SED properties, we suggested that I05399 is in a transition 
phase from Class 0 to I, whereas it was previously classified as a Class I 
object (e.g., \cite{bally2002} and references therein). 
Miettinen et al. (2010; hereafter, Paper II) performed  
follow-up NH$_3$ observations of the Orion B9 cores. The gas kinetic 
temperature and one-dimensional non-thermal velocity dispersion towards 
I05399 and SMM 1 were derived to be $13.5\pm1.6$ and $11.9\pm0.9$ K, and 
0.24 and 0.27 km~s$^{-1}$, respectively. The latter values are transonic, 
i.e., $\sigma_{\rm NT}\sim c_{\rm s}$, where $c_{\rm s}$ is the isothermal 
sound speed. We note that the spatial resolution of our NH$_3$ 
observations, $40\arcsec$ or 0.09 pc at 450 pc, corresponds to the typical size
of dense cores and the sonic scale length of the interstellar medium, 
$\lambda_{\rm s}\sim0.1$ pc, where turbulence transitions from supersonic to 
subsonic (\cite{vazquez2003}). Therefore, at the sonic scale we expect
$\sigma_{\rm NT}\simeq c_{\rm s}$. 

Employing the new temperature values with the assumption that 
$T_{\rm kin}=T_{\rm dust}$, the masses of I05399 and SMM 1 were estimated to be 
$6.1\pm1.4$ and $8.4\pm1.5$ M$_{\odot}$. 
The virial-parameter analysis of SMM 1 suggested that it is 
gravitationally bound, supporting our earlier speculation that it is 
in the prestellar phase of evolution. 
Miettinen et al. (2012; hereafter, Paper III) presented further 
molecular-line observations towards I05399 and SMM 1. The CO depletion 
factors towards these sources were estimated to be $3.2\pm0.7$ and 
$1.9\pm0.3$, respectively. Despite the fact that no evidence of a 
significant CO depletion in SMM 1 was found, we derived a very high level of 
N$_2$D$^+$/N$_2$H$^+$ deuteration, namely $0.992\pm0.267$. On the other hand, 
towards I05399, the DCO$^+$/HCO$^+$ and N$_2$D$^+$/N$_2$H$^+$ 
column density ratios were found to be $0.020\pm0.004$ and $0.207\pm0.046$. 
A lower limit to the fractional ionisation of $x({\rm e})>1.5\times10^{-8}$ 
was derived in I05399, which is over six times lower than the 
corresponding value determined by Bergin et al. (1999). 
The cosmic-ray ionisation rate of H$_2$ was estimated to be 
$\zeta_{\rm H_2}\sim2.6\times10^{-17}$ s$^{-1}$, which is within a factor of two 
of the standard value $1.3\times10^{-17}$ s$^{-1}$ (e.g., \cite{caselli1998}). 
We also detected D$_2$CO emission from both I05399 and SMM 1. 
We note that D$_2$CO, a molecule expected to form on the surface 
of dust grains, was also detected in starless/prestellar cores by 
Bacmann et al. (2003) and Bergman et al. (2011). Its presence in the gas phase 
requires an efficient desorption mechanism, which in starless cores could be 
due to cosmic-ray bombardment and/or the release of the formation energy of 
the newly formed species (see Paper III and references therein). 
However, in SMM 1 the gas-phase D$_2$CO could originate in shocks driven by 
the HH 92 jet from I05399. This conforms to the low CO depletion factor 
derived towards SMM 1. Overall, the results of Paper III suggest very peculiar 
chemical properties in the I05399/SMM 1 -system.

Recent \textit{Herschel} data show that I05399/SMM 1 is embedded in a  
northeast-southwest oriented filamentary structure (see \cite{miettinen2012}; 
Fig.~9 therein). Miettinen (2012) found that there is a sharp
velocity gradient in the region (across the short axis of the parent 
filament), and speculated that it might represent a 
shock front resulting from the feedback from the nearby expanding \ion{H}{ii} 
region NGC 2024. As for the other members of Orion B9, the formation of the 
I05399/SMM 1 -system might have been triggered by this feedback.
Finally, we note that I05399 is the source No.~3133 in the 
\textit{Spitzer} YSO survey of Orion molecular clouds (\cite{megeath2012}), 
and No.~310 in the \textit{Herschel} spectroscopic survey of protostars in 
Orion (\cite{manoj2013}). Manoj et al. (2013) observed far-infrared CO lines, 
and suggested that the emission originates in low-density gas 
[$n({\rm H}_2)=6.3^{+9.5}_{-6.2}\times10^3$ cm$^{-3}$] heated by 
outflow shocks ($T=3\,548^{+918}_{-729}$ K) near the protostar (at radii 
$\gtrsim$ several 100--1\,000 AU).

To broaden our view of the I05399/SMM 1 -system and its physical properties, we 
mapped it in the 350-$\mu$m dust conti\-nuum emission with the 
Submillimetre APEX Bolometer Camera (SABOCA) on APEX. This allowed us to 
improve the angular resolution by a factor of about 2.6 compared to our 
previous LABOCA map of the source. In this paper, we discuss the results of 
our new high-resolution 350-$\mu$m data, and use it in conjuction 
with the LABOCA 870-$\mu$m data. The observations, data, and data reduction 
are described in Sect.~2. We present the observational results in Sect.~3, 
whereas analysis and its results are presented in Sect.~4. We discuss the 
results in Sect.~5 and in Sect.~6, we summarise the paper and draw our main 
conclusions.

\section{Observations and data reduction}

\subsection{SABOCA 350-$\mu$m imaging}

A field of $7\farcm0 \times 6\farcm1$ (0.012 deg$^2$), centred between I05399 
and SMM 1 on the coordinates $\alpha_{2000.0}=05^{\rm h}42^{\rm m}29\fs6$, 
$\delta_{2000.0}=-01\degr 20\arcmin 19\farcs2$, was mapped with SABOCA 
(\cite{siringo2010}) on the APEX 12-m telescope (\cite{gusten2006}) at Llano de 
Chajnantor in the Atacama desert of the Chilean Andes. 
This site provides very good weather conditions for submm astro\-nomy (e.g., 
\cite{tremblin2012}). SABOCA is a 37-channel on-sky TES 
(transition edge sensor) bolometer array operating at
350 $\mu$m, with a nominal resolution of $7\farcs7$ (HPBW). The effective
field of view of the array is $1\farcm5$. The SABOCA passband
has an equi\-valent width of about 120 GHz centred on an effective frequency 
of 852 GHz. The observations were carried out on 22 and 23 August 2012. 
The atmospheric attenuation at 350 $\mu$m was monitored using the sky-dip 
method, and the zenith opacity was found to be in the range 
$\tau_{\rm z}^{350\,\mu {\rm m}}=0.84-1.17$. The amount of precipitable water 
vapour (PWV) was $\sim0.3-0.5$ mm. The telescope focus and pointing were 
optimised and checked at re\-gular intervals on the pla\-nets Mars and Uranus, 
the red supergiant VY Canis Majoris (VY CMa), FU Orionis star V883 Ori, 
infrared cluster NGC 2071 IR, and the massive YSO/H$_2$O maser source 
G305.80-0.24 (B13134). The absolute calibration 
uncertainty for SABOCA is $\sim30\%$. Mapping was performed using a 
``fast-scanning'' method without chopping the secondary mirror 
(\cite{reichertz2001}). The mosaic was constructed by combining 30 individual 
fast-scanning ($\sim1\farcm2-1\farcm5$~s$^{-1}$) maps. The total on-source 
integration time was 4.2 h (during the UTC time ranges 08:42--14:02 and 
15:26--16:39).

Data reduction was done with the pipeline iterations of the CRUSH-2 
(Comprehensive Reduction Utility for SHARC-2) (version 2.12-2) software 
package\footnote{{\tt http://www.submm.caltech.edu/$\sim$sharc/crush/index.htm}} (\cite{kovacs2008}). During the reduction process, smoothing with a Gaussian 
kernel of the size $1\farcs41$ was applied. The instrument beam FWHM used in 
CRUSH-2 is $7\farcs5$, so the angular resolution of the final image is 
$7\farcs63$ (0.017 pc or 3\,433.5 AU at 450 pc). 
The gridding was performed with a cell size of $1\farcs5$. The resulting 
$1\sigma$ rms noise level in the final map is $\sim80$ mJy~beam$^{-1}$. 

\subsection{LABOCA 870-$\mu$m data}

Our LABOCA (\cite{siringo2009}) 870-$\mu$m observations were already published 
in Paper I. The data was originally reduced using the BOlometer array data 
Analysis software (BoA; \cite{schuller2012}). 
However, to eliminate the effect of data-reduction 
software, and to be properly comparable with our present SABOCA data, we 
re-reduced the LABOCA data with CRUSH-2. Smoothing with a Gaussian 
kernel of the size $3\farcs76$ was applied, while the instrument beam FWHM 
used in CRUSH-2 is $19\farcs5$. Therefore, the angular resolution of the final 
image is $19\farcs86$ (0.043 pc at 450 pc). The gridding was performed with a 
cell size of $4\farcs0$. The resulting $1\sigma$ rms noise level in the final 
map is 30 mJy~beam$^{-1}$. For comparison, in Paper I the resolution and rms 
noise were $20\farcs13$ and 30 mJy~beam$^{-1}$. The resulting peak surface 
brightnesses of the sources are very similar to the values in the BoA-reduced 
map.

In Paper III, we noticed that our LABOCA map employed in Papers I and II was 
misaligned, and therefore we had to adjust its pointing (see footnote~2 in 
Paper III). The target positions of our previous molecular-line observations 
were chosen to be the peak positions of the LABOCA map before adjusting the 
pointing, and therefore they are slightly offset from the 870-$\mu$m maxima. 
The misalignment of the map appears to be the result of our data reduction 
with BoA. In the present work, where the LABOCA data is re-reduced with 
CRUSH-2, there is a good agreement between the coordinates of the SABOCA and 
LABOCA peak positions (Sect.~3.2). 

\subsection{Archival data from the Spitzer Space Telescope}

In this study, we also employ the infrared data from \textit{Spitzer} 
(\cite{werner2004}) to characterise the star-formation activity in the 
I05399/SMM 1 -system. The Spitzer Science Center 
(SSC)\footnote{{\tt http://ssc.spitzer.caltech.edu/}} and Infrared 
Science Archive (IRSA)\footnote{{\tt http://irsa.ipac.caltech.edu/}} recently 
announced the release of a set of Enhanced Imaging Products (SEIP) from the 
Spitzer Heritage Archive 
(SHA)\footnote{{\tt http://sha.ipac.caltech.edu/applications/Spitzer/SHA/}}. 
These include Super Mosaics and a Source List of photometry for compact 
sources. 

The IRAC (Infrared Array Camera) Super Mosaics were downloaded from SHA. 
The IRAC instrument took simultaneous images at wavelengths of 3.6, 4.5, 
5.8, and 8.0 $\mu$m with an angular resolution between $1\farcs5$ and 
$1\farcs9$ (\cite{fazio2004}). We also retrieved the 24- and 70-$\mu$m MIPS 
(Multiband Imaging Photometer for Spitzer; \cite{rieke2004}) images towards 
I05399/SMM 1. The angular resolution at these wavelengths is $6\arcsec$ and 
$18\arcsec$, respectively. The MIPS data are part of the programme ``A MIPS 
Survey of the Orion L1641 and L1630 Molecular Clouds'' (PI: G. Fazio; 
programme name/ID: ORION$_{-}$MIPS/47; AOR-Key: 12647424).

I05399 is designated as SSTSLP J054227.67-012000.4 in the Source List. 
In the $3\farcs8$ diameter aperture, the reported IRAC flux densities in mJy 
are $1.317\pm0.005$ (3.6 $\mu$m), $5.915\pm0.001$ (4.5 $\mu$m; 
bandfill value), $7.195\pm0.006$ (5.8 $\mu$m; bandfill), and 
$9.796\pm0.018$ (8.0 $\mu$m). The 24-$\mu$m flux density, from a PSF 
fitting method, is $1.153\pm0.524$ Jy. The latter is close to the value 
$1.3\pm0.05$ Jy determined in Paper I.

\section{Observational results}

\subsection{Images}

\begin{figure*}
\begin{center}
\includegraphics[width=0.48\textwidth]{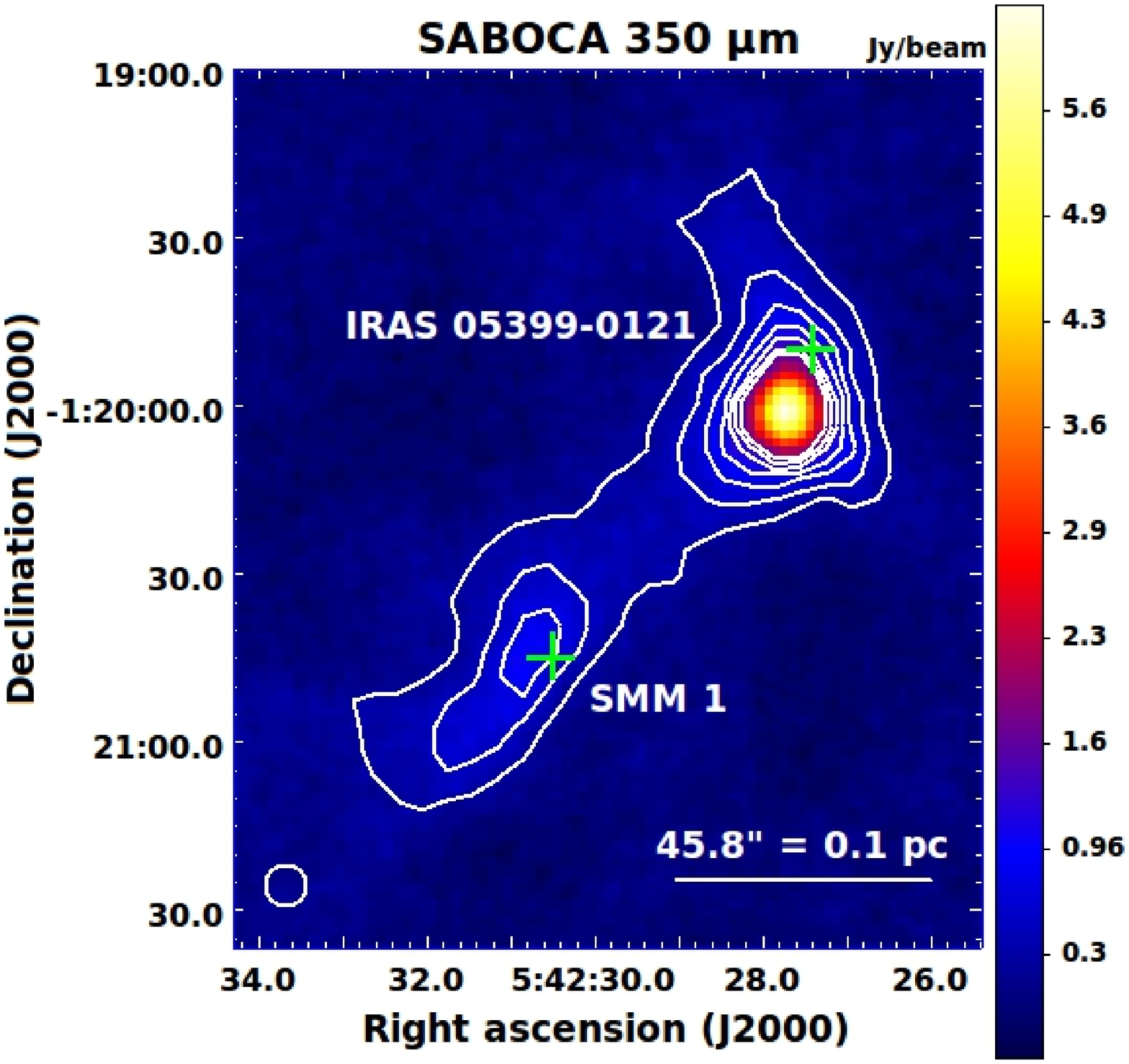}
\includegraphics[width=0.483\textwidth]{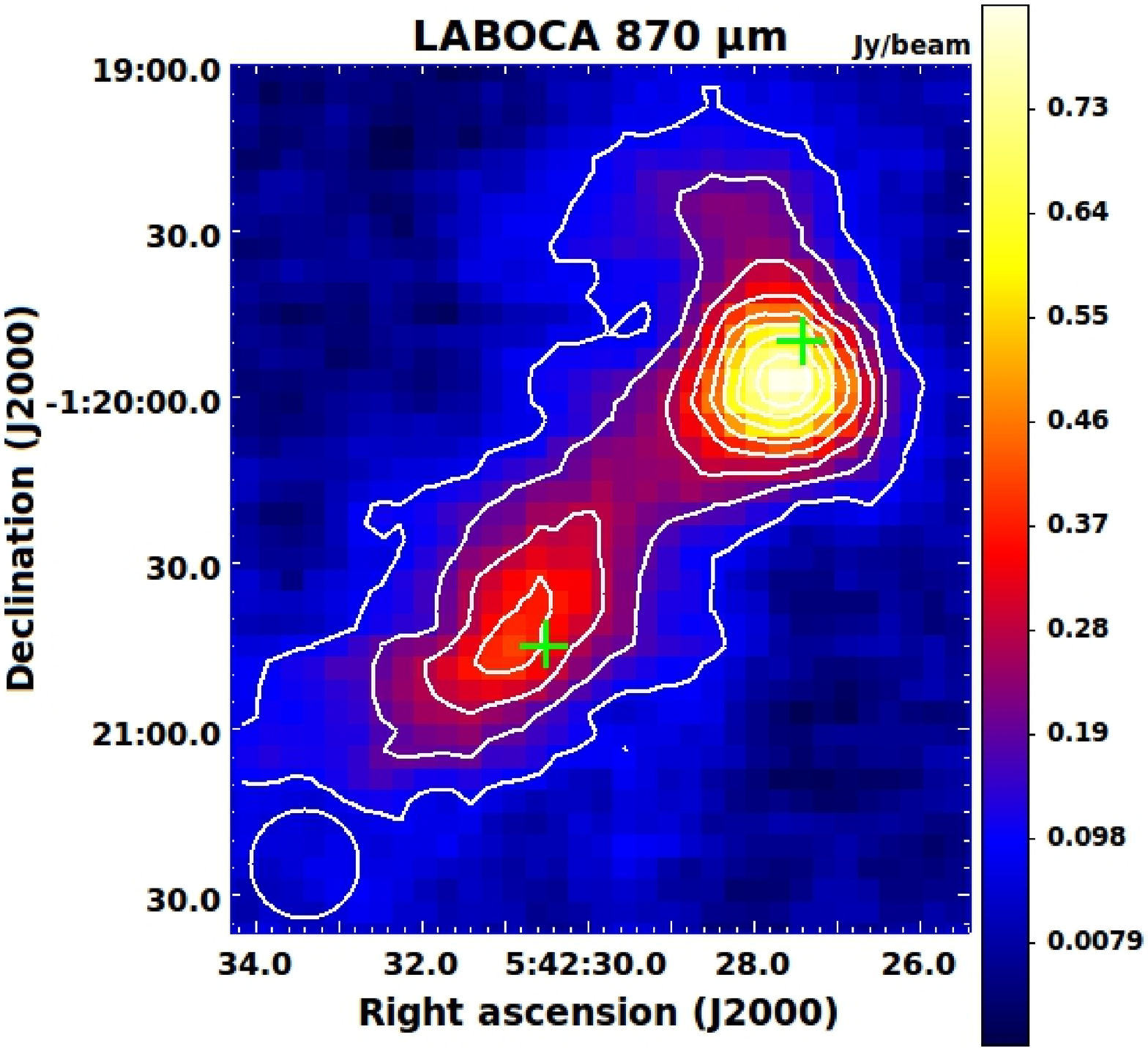}
\caption{SABOCA 350-$\mu$m and LABOCA 870-$\mu$m images of the I05399/SMM 1 
-system. The images are shown with linear scaling, and the colour bars 
indicate the surface-brightness scale in Jy~beam$^{-1}$. The overlaid 
SABOCA contours start at three times the noise level ($3\sigma=240$ 
mJy~beam$^{-1}$) and increase at this interval to 1\,680 mJy~beam$^{-1}$. 
The LABOCA contour levels range from 90 mJy~beam$^{-1}$ ($3\sigma$) to 720 
mJy~beam$^{-1}$, increasing in steps of $3\sigma$. The green plus signs show 
the target positions of our previous molecular-line observations. 
A scale bar indicating the 0.1 pc projected length is shown in the left panel, 
with the assumption of a 450 pc line-of-sight distance. The effective beam 
sizes ($7\farcs63$ for SABOCA, $19\farcs86$ for LABOCA) are shown in the lower 
left corner of the panels.}
\label{figure:maps}
\end{center}
\end{figure*}

The SABOCA and LABOCA images are shown in Fig.~\ref{figure:maps}. The overall 
filamentary emission morphology seen in the maps is rather similar to each 
other. The projected length along the long axis of the 350-$\mu$m 
filament is about $2\arcmin$ or 0.26 pc. The projected average width of this 
filament is about $\sim30\arcsec$ or $\sim0.07$ pc. Its aspect ratio is 
therefore roughly $A \simeq4$. The projected separation between the 350-$\mu$m 
peaks of I05399 and SMM 1 is $\sim1\farcm1$ or 0.14 pc.

\begin{figure*}
\begin{center}
\includegraphics[width=0.49\textwidth]{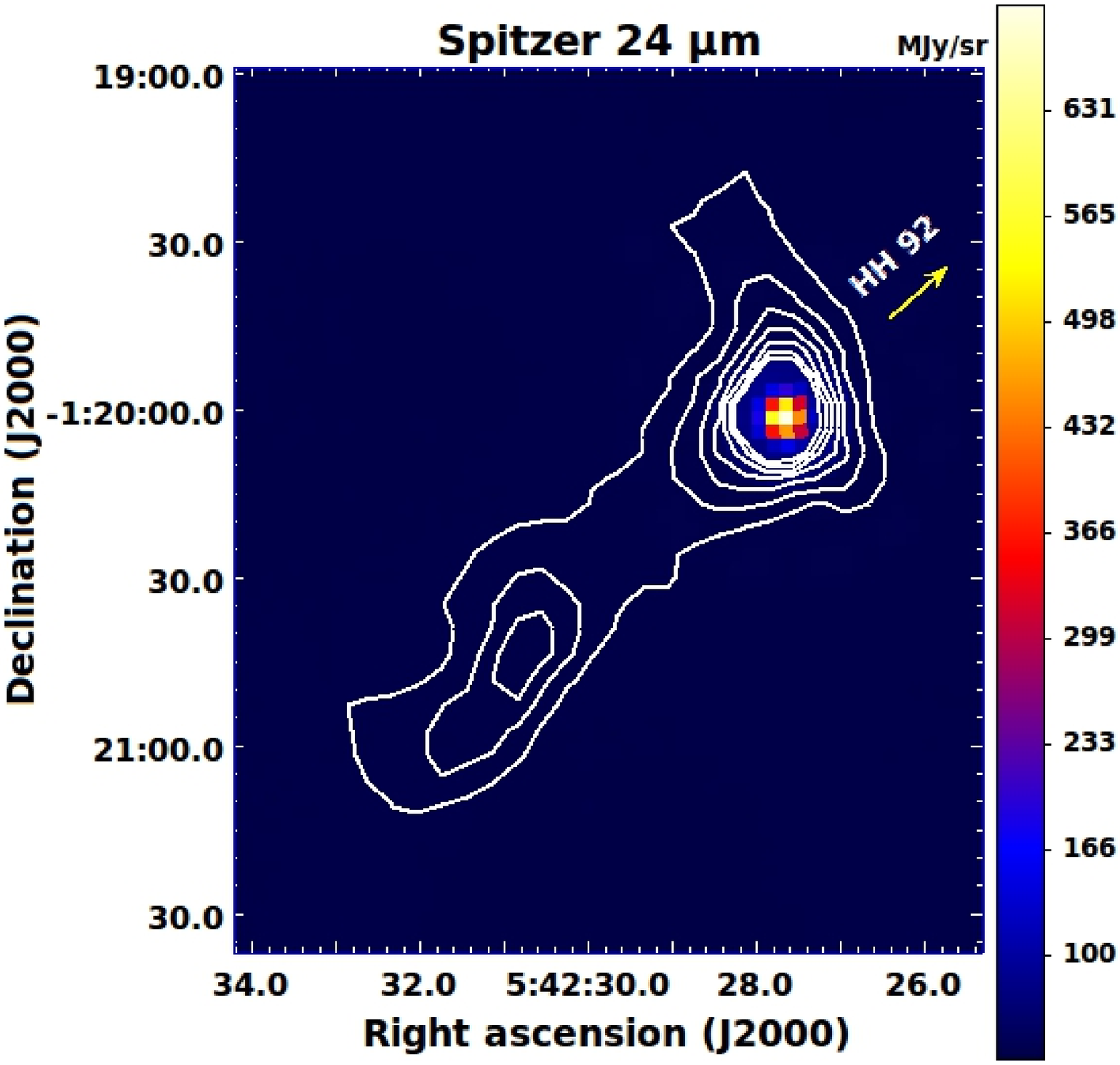}
\includegraphics[width=0.49\textwidth]{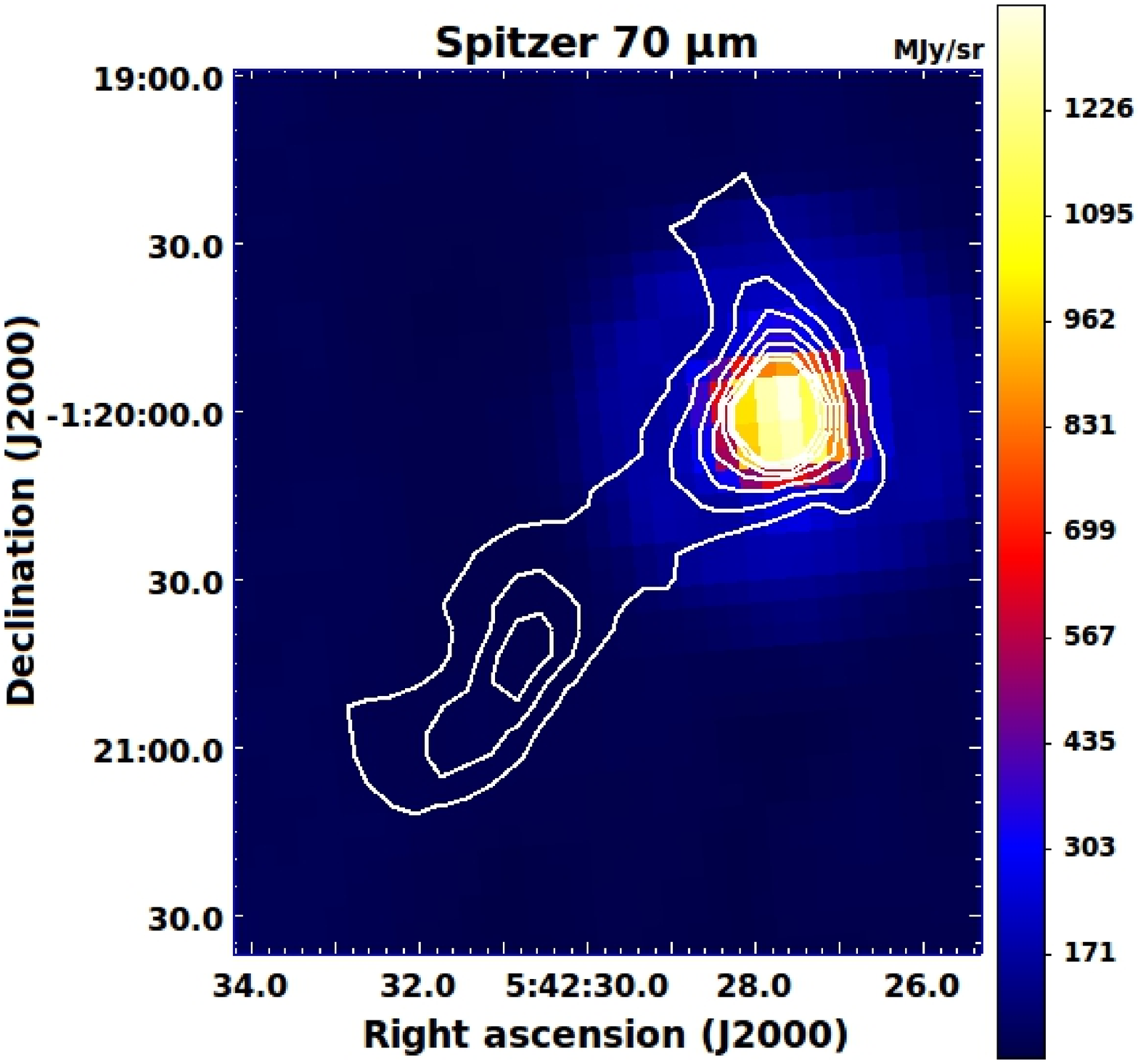}
\includegraphics[width=0.7\textwidth]{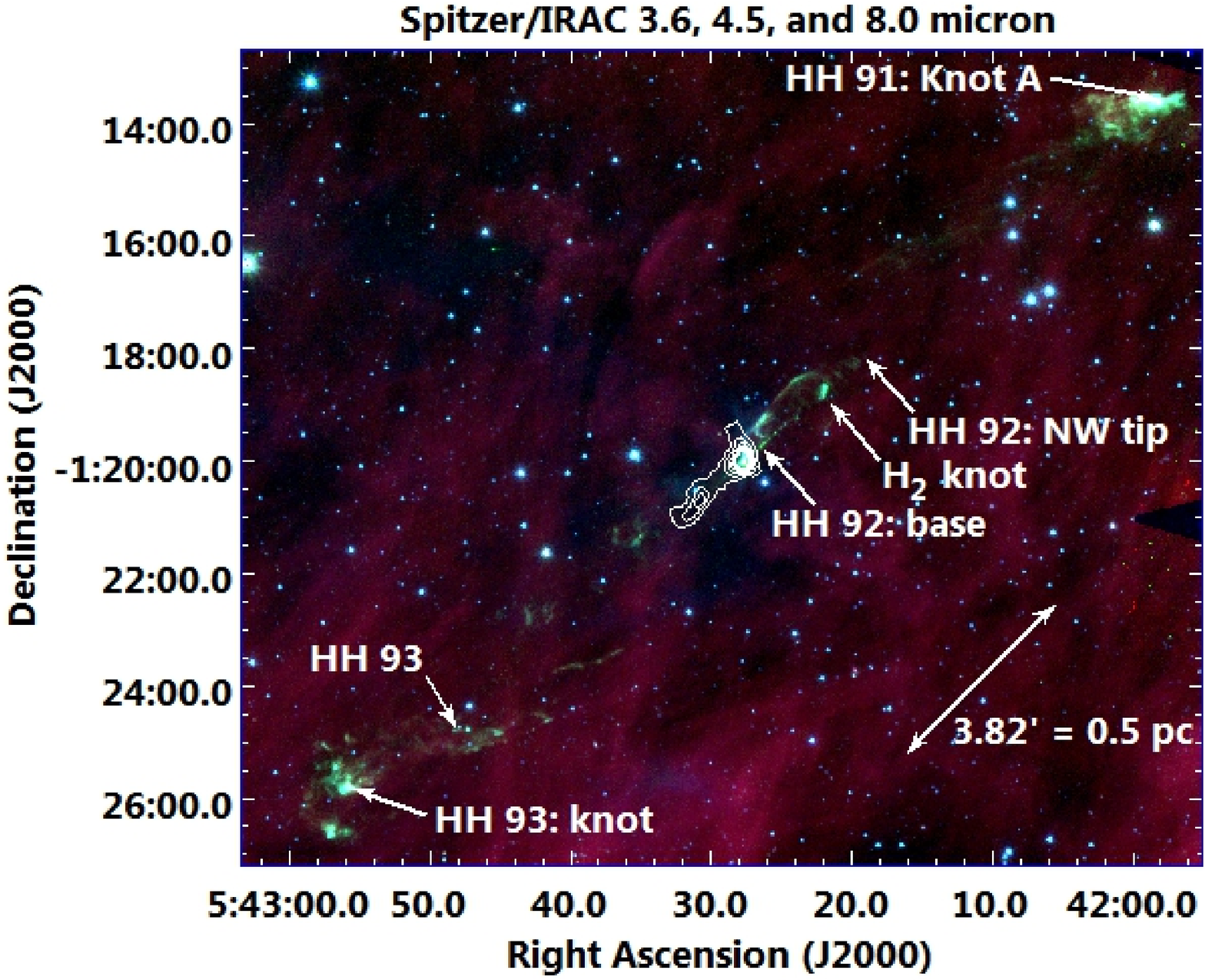}
\caption{\textbf{Top panels:} \textit{Spitzer}/MIPS 24- and 70-$\mu$m images 
of the I05399/SMM 1 -system. The images are shown with linear scaling, and the 
colour bars indicate the surface-brightness scale in MJy~sr$^{-1}$. The images 
are overlaid with contours of SABOCA 350-$\mu$m dust continuum emission as in 
Fig.~\ref{figure:maps}. In the left panel, the yellow arrow starts from the 
base of the HH 92 jet driven by I05399 (\cite{bally2002}). 
\textbf{Bottom panel:} \textit{Spitzer}/IRAC three-colour composite image 
showing the I05399/SMM 1 -system in the centre and part of the outflow features 
associated with the HH 92 jet. The 3.6, 4.5, and 8.0 $\mu$m emission is coded 
in blue, green, and red, respectively, and the colours are shown in 
logarithmic scales. The contours show the 350-$\mu$m emission as in the top 
panels. A scale bar indicating the 0.5 pc projected length is indicated. Note 
that the total length of the outflow, $34\arcmin$, corresponds to 4.45 pc at 
450 pc.}
\label{figure:spitzer}
\end{center}
\end{figure*}

The \textit{Spitzer} images are shown in Figs.~\ref{figure:spitzer} and 
\ref{figure:jet}. As can be seen from the images, I05399 is associated with 
24- and 70-$\mu$m point sources, whereas SMM 1 shows no such emission. 
As described in Sect.~1, I05399 drives the HH 92 protostellar jet. 
In the \textit{Spitzer}/IRAC three-colour image, we have labelled several 
of the outflow features associated with HH 92 (see \cite{bally2002}; 
Table~1 therein). The very long length of the outflow could mean that we are 
viewing I05399 from an edge-on orientation so that the outflow lies in the 
plane of the sky. The 4.5-$\mu$m band is particularly useful for studying 
protostellar jets as it is sensitive to shock-excited line features 
(H$_2$ and CO; see, e.g., \cite{smith2005}; \cite{ybarra2009}; 
\cite{debuizer2010}). From the base to the northwest tip of the HH 92 jet, 
a chain of several 4.5-$\mu$m knots are 
apparent. The brightest H$_2$ knot reported by Bally et al. (2002) is also 
associated with a strong 4.5-$\mu$m feature. These knots might be the result 
of internal shocks within the jet. The knot HH 91A, showing extended 
4.5-$\mu$m emission, is a large shock front, and the 4.5-$\mu$m features of 
HH 93 show other shocks associated with the counterflow (in 
Fig.~\ref{figure:spitzer} we have labelled the closest point to I05399 and the 
brightest knot of HH 93). A visual inspection of the 3.6- and 
4.5-$\mu$m images provides hint of a quadrupolar outflow, supporting the 
suggestion that I05399 is actually a binary system.

\subsection{The submm cores}

To extract the 350- and 870-$\mu$m core properties, we employed the commonly 
used two-dimensional Clumpfind algorithm, {\tt clfind2d}, developed by 
Williams et al. (1994). The algorithm requires two configuration
parameters: \textit{i)} the intensity threshold, i.e., the lowest contour
level, which determines the minimum emission to be included
into the core; and \textit{ii)} the contour level spacing, which
determines the required ``contrast'' between two cores to be
considered as different objects. We set both of these parameters
to $3\sigma$ or 240 mJy~beam$^{-1}$ for SABOCA and 90 mJy~beam$^{-1}$ for 
LABOCA. Blended source structures are assigned via a 
friends-of-friends algorithm to the closest emission peak. The stepsize 
used for contouring the data can affect the total number of identified 
structures: if contour spacings are too large real fragments can be missed, 
while closely spaced contours can cause artificial noise features to be 
identified as ``real'' objects (\cite{pineda2009}). However, our source is 
relatively simple and the Clumpfind results are therefore expected to be 
robust. Moreover, Pineda et al. (2009) concluded that that the 2D version of 
Clumpfind used here is more reliable than the three-dimensional version.

The J2000.0 coordinates of the peak emission, angular and linear 
core effective radii ($R_{\rm eff}=\sqrt{A_{\rm proj}/\pi}$, where 
$A_{\rm proj}$ is the projected area within the $3\sigma$ contour), peak 
surface brightnesses, and integrated flux densities (within $3\sigma$) are 
listed in Cols.~(3)--(8) of Table~\ref{table:cores}. The effective 
radii listed in Table~\ref{table:cores} are not corrected for beam size.
The quoted flux density uncertainties were derived from the rms noise and the 
absolute calibration error (30\% for SABOCA and 10\% for LABOCA), which were 
combined quadratically. We would like to point out that the LABOCA emission 
peak of I05399 lies $8\farcs8$ southeast of the previously determined position 
(in the BoA-reduced map), whereas for SMM 1 the revised maximum lies 
$6\arcsec$ to the east. 

\begin{table*}
\caption{SABOCA 350- and LABOCA 870-$\mu$m characteristics of I05399 and 
SMM 1.}
\begin{minipage}{2\columnwidth}
\centering
\renewcommand{\footnoterule}{}
\label{table:cores}
\begin{tabular}{c c c c c c c c}
\hline\hline 
Source & $\lambda$ & $\alpha_{2000.0}$ & $\delta_{2000.0}$ & \multicolumn{2}{c}{$R_{\rm eff}$} & $I_{\nu}^{\rm peak}$ & $S_{\nu}$ \\

       & [$\mu$m] & [h:m:s] & [$\degr$:$\arcmin$:$\arcsec$] & [\arcsec] & [pc] & [Jy~beam$^{-1}$] & [Jy] \\
\hline
IRAS 05399-0121 & 350 & 05 42 27.7 & -01 20 00.4 & 19.3 & 0.042 & $6.24\pm1.87$ & $20.66\pm6.20$ \\
                & 870 & 05 42 27.7 & -01 19 57.0 & 35.3 & 0.077 & $0.82\pm0.09$ & $2.09\pm0.21$ \\
SMM 1 & 350 & 05 42 30.9 & -01 20 45.4 & 18.7 & 0.041 & $1.03\pm0.32$ & $7.96\pm2.39$\\
      & 870 & 05 42 30.9 & -01 20 45.0 & 33.1 & 0.072 & $0.40\pm0.05$ & $1.49\pm0.15$\\
\hline 
\end{tabular} 
\end{minipage}
\end{table*}

\section{Analysis and results}

\subsection{Dust temperature and H$_2$ column density maps}

The present 350-$\mu$m data together with our previous 870-$\mu$m data allow 
us to determine the dust colour temperature ($T_{\rm dust}$) and H$_2$ column 
density [$N({\rm H_2})$] distributions in the I05399/SMM 1 -system.

We smoothed the SABOCA map to the resolution of our LABOCA data 
($7\farcs63 \rightarrow 19\farcs86$), and regridded the maps to the same pixel 
size ($4\arcsec$). Convolving the SABOCA map halved the $1\sigma$ rms noise 
level to 40 mJy~beam$^{-1}$. The intensity ratio map was then converted into a 
$T_{\rm dust}$ map with the dust emissivity index fixed to $\beta=1.8$. The 
choice of this value of $\beta$, which lies between the usual 
fiducial values of $\beta=1-2$, is explained below. Fitting 
$\beta$ would require more than two photometric bands, particularly on the 
Rayleigh-Jeans side of the SED (e.g., \cite{doty1994}; \cite{shetty2009}). 
Unfortunately, the SCUBA Legacy 450-$\mu$m map does not cover the I05399/SMM 1 
-system completely (about half of SMM 1 lies outside the map's boundaries). 
The Orion B9 region was observed as part of the ``\textit{Herschel} 
Gould Belt Survey (HGBS)'' (\cite{andre2010})\footnote{\textit{Herschel} is an 
ESA space observatory with science instruments provided by European-led 
Principal Investigator consortia and with important participation from NASA 
(\cite{pilbratt2010}). The HGBS is a \textit{Herschel} Key Programme jointly 
carried out by SPIRE Specialist Astronomy Group 3 (SAG 3), scientists of 
several institutes in the PACS Consortium (CEA Saclay, INAF-IFSI Rome and 
INAF-Arcetri, KU Leuven, MPIA Heidelberg), and scientists of the 
\textit{Herschel} Science Center (HSC). For more details, see 
{\tt http://gouldbelt-herschel.cea.fr}. At the moment, the unpublished 
Level-2 map products are available in the \textit{Herschel} Science Archive 
(HSA)\footnote{{\tt http://herschel.esac.esa.int/Science$_{-}$Archive.shtml}} 
but these data do not have an absolute intensity level (M.~Hennemann, private 
communication). Unlike \textit{Herschel}, the ground-based bolometers SABOCA 
and LABOCA heavily filter the emission on large scales in order to remove 
the atmospheric emission. Because of the current quality of the 
\textit{Herschel} data on I05399/SMM 1, we do not employ them in the present 
study.} 
%whereas employing all the \textit{Herschel}/SPIRE bands (up to 500 $\mu$m) 
%would force us to sacrifice the present angular resolution (\textit{Herschel} 
%beam FWHM at 500 $\mu$m is $36\arcsec$). 
Besides the value of $\beta$ and the paucity of wavelength bands 
used, another important factor affecting the derived dust colour temperatures 
is that our 870-$\mu$m continuum data could be significantly contaminated by 
the $^{12}$CO$(3-2)$ line emission at 345 GHz (\cite{drabek2012}). 
Although CO does not appear to be significantly depleted 
in the studied source (\cite{bergin1999}; Paper III), the CO contamination is 
not expected to be more than a few tens of percent. Mapping observations of 
CO$(3-2)$ would be required to quantify this effect. The constructed map of 
line-of-sight-averaged dust temperature is shown in the left panel of 
Fig.~\ref{figure:dust}. 

As expected, I05399 appears as a warm spot with $T_{\rm dust}$ values up to 
$\sim50$ K in the temperature map due to embedded heating source. 
A higher $\beta$ value would lower the temperature for the same flux 
density ratio. Within the aperture with size equal to the effective 
350-$\mu$m angular size ($R_{\rm eff}=19\farcs3$), and centred on the SABOCA 
peak, the average $T_{\rm dust}$ towards I05399 is $22.3\pm10.2$ K, where the 
$\pm$-error represents the standard deviation (hereafter referred to as std). 
This is comparable to the value $\sim18.5$ K obtained from the SED fit in 
Paper I. On the other hand, the mean and std of $T_{\rm dust}$ is $15.3\pm7.0$ K 
in a $40\arcsec$-diameter aperture centred on the position of our previous 
NH$_3$ measurements with $40\arcsec$ resolution (Paper II). This is close to 
the measured gas temperature of $13.5\pm1.6$ K. As can be seen in 
Fig.~\ref{figure:dust}, the temperature is colder in other parts of the 
system. Measuring $T_{\rm dust}$ within $R_{\rm eff}=18\farcs7$ and centred on 
the SABOCA peak of SMM 1, the mean$\pm$std is $12.5\pm1.2$ K. 
Towards our line observation position, the corresponding value is $11.4\pm3.0$ 
K within a $40\arcsec$ aperture. This is strikingly similar to the $T_{\rm kin}$ 
value of $11.9\pm0.9$ K derived from ammonia. The $T_{\rm dust}$ rises to 
$\sim15-16$ K at the southeastern edge of the source.

The $N({\rm H_2})$ map was constructed as follows. We first computed the 
optical thickness map at 870 $\mu$m. The $\tau_{{\rm 870\, \mu m}}$ map was 
then divided by the 870-$\mu$m dust extinction cross-section per H$_2$ 
molecule [cf.~Eq.~(9) in Miettinen \& Harju (2010)]. We adopted the 
solar abundances for hydrogen, helium, and heavier elements, i.e., 
their mass fractions were taken to be 0.71, 0.27, and 0.02 of the total mass 
of all elements, respectively. This corresponds to a helium to hydrogen 
abundance ratio of about 0.10 and mean molecular weight per H$_2$ molecule of
$\mu_{\rm H_2}\simeq2.82$ (\cite{kauffmann2008}; Appendix~A.1 therein). 
The 870-$\mu$m dust mass absorption (or emission) coefficient, i.e., dust 
opacity per unit dust mass was taken to be 1.38 cm$^2$~g$^{-1}$. 
This value was interpolated from the widely used Ossenkopf \& Henning (1994, 
hereafter OH94) model describing graphite-silicate dust grains that have 
coagulated and accreted \textit{thin}\footnote{In Papers I--III, we assumed 
that the grains have thick ice mantles, and used the 870-$\mu$m opacity of 
$\kappa_{{\rm 870\, \mu m}}\simeq1.7$ cm$^2$~g$^{-1}$ from OH94. The 
assumption of thin ice mantles is supported by the fact that no significant 
CO depletion was found in Paper III.} ice mantles over a period of $10^5$ yr 
at a gas density of 
$n_{\rm H}= n({\rm H}) + 2n({\rm H_2})\simeq 2n({\rm H_2}) = 10^5$ cm$^{-3}$. 
For the average dust-to-hydrogen mass ratio, $M_{\rm dust}/M_{\rm H}$, we 
adopted the canonical value 1/100 (e.g., \cite{draine2011}; 
Table~23.1 therein), although the true value can be different 
(e.g., \cite{vuong2003}; \cite{draine2007}). The total dust-to-gas mass ratio 
used is therefore $M_{\rm dust}/M_{\rm gas}=M_{\rm dust}/(1.41M_{\rm H})=1/141$, 
where the factor 1.41 ($\simeq1/0.71$) is the ratio of total mass 
(H+He+metals) to hydrogen mass. In the adopted OH94 dust model, 
$\beta \simeq1.8$, as determined from the slope between 350 and 
1\,300 $\mu$m ($\kappa_{\lambda}\propto \lambda^{-\beta}$). Therefore, this value 
was adopted in the derivation of the $T_{\rm dust}$ map 
to be consistent with the dust properties used. The resulting column 
density map is shown in the right panel of Fig.~\ref{figure:dust}. Both I05399 
and SMM 1 stand out as dense portions of the filamentary structure. The 
$N({\rm H_2})$ values towards these two cores are $\sim 5-6\times10^{22}$ 
and $\sim 3-4\times10^{22}$ cm$^{-2}$, respectively.

\begin{figure}[!h]
\centering
\resizebox{0.98\hsize}{!}{\includegraphics{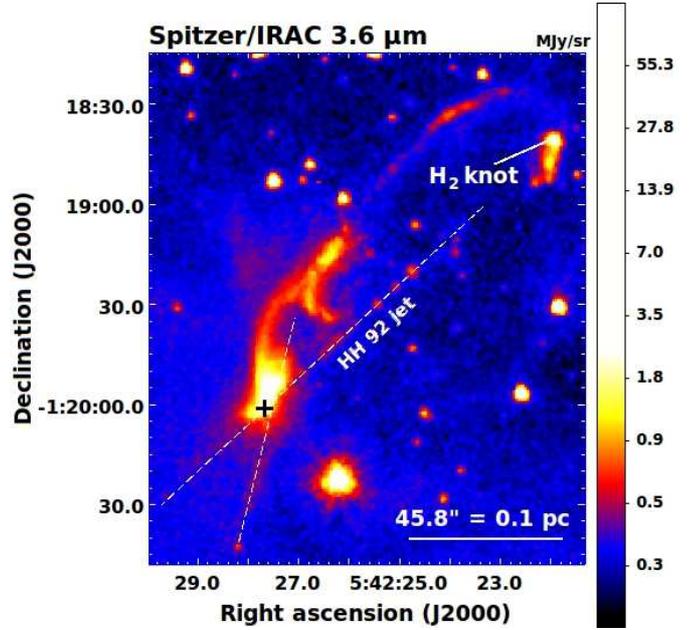}}
\caption{A zoomed in view of Fig.~\ref{figure:spitzer} to show the 
\textit{Spitzer}/IRAC 3.6-$\mu$m emission towards I05399. The image is 
displayed with logarithmic scaling to highlight the jet features. 
The dashed lines illustrate the possible quadrupolar outflow structure.
Note that SMM 1 lies outside the image. The black plus sign shows the peak of 
the 24-$\mu$m source.}
\label{figure:jet}
\end{figure}

\begin{figure*}
\begin{center}
\includegraphics[width=0.48\textwidth]{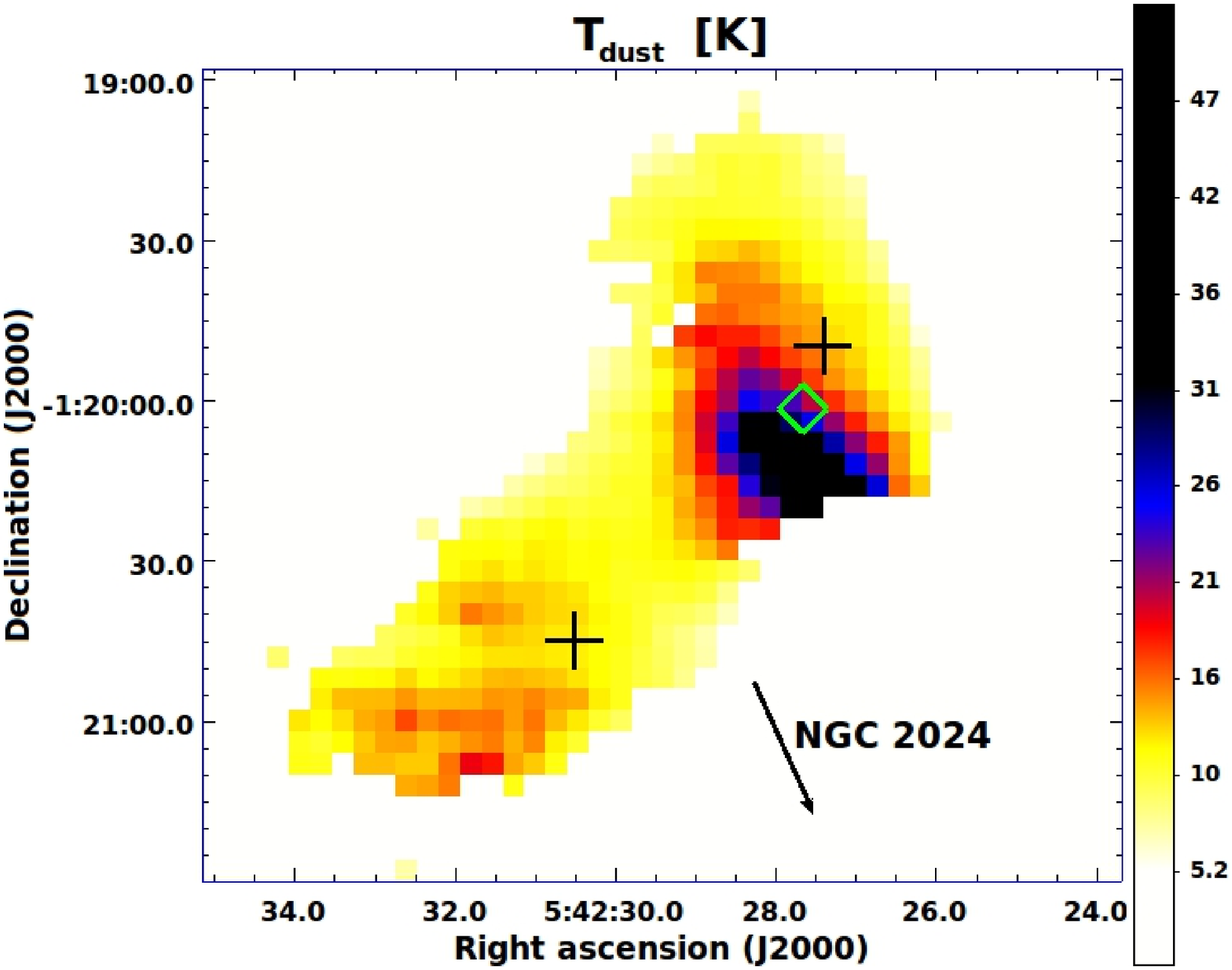}
\includegraphics[width=0.516\textwidth]{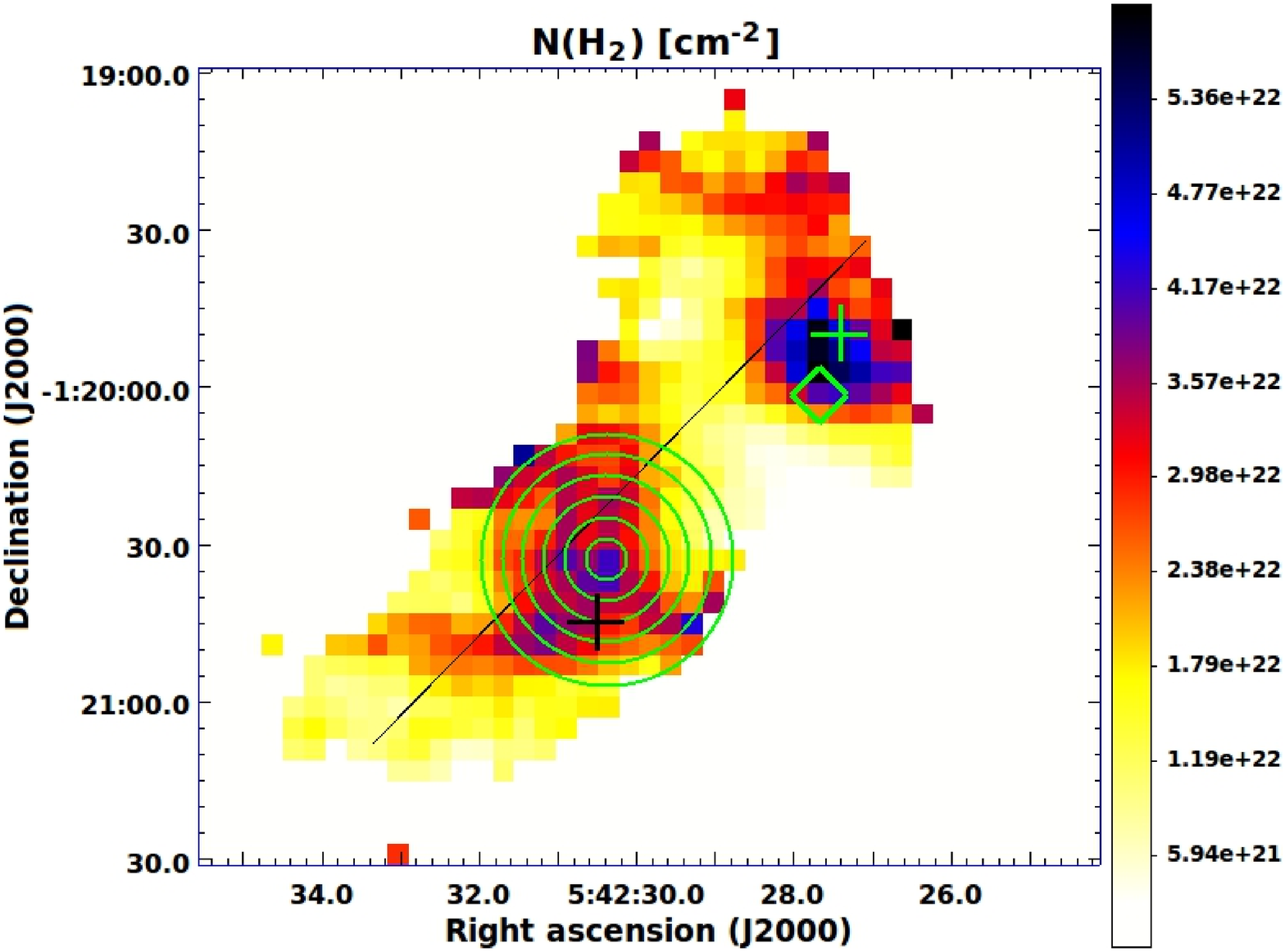}
\caption{Distributions of dust colour temperature (\textit{left}) and H$_2$ 
column density (\textit{right}). The dust emissivity index, $\beta$, was 
fixed at 1.8. The colour-scale bars on the right indicate the units in K and 
cm$^{-2}$, respectively. The plus signs show the target positions of 
our previous molecular-line observations. The diamond symbol indicates the 
24-$\mu$m emission peak of I05399. The direction towards NGC 2024 is indicated 
by an arrow in the left panel. The annuli used to compute the density profile 
shown in Fig.~\ref{figure:profile} are shown by the green circles, 
while the black solid line in the right panel shows the location of the slice 
used to extract the central column density discussed in Sect.~4.5. 
The resolution of the images is $19\farcs86$ or about 0.043 pc.}
\label{figure:dust}
\end{center}
\end{figure*}

\subsection{Distribution of the dust emissivity index}

\begin{figure}[!h]
\centering
\resizebox{0.95\hsize}{!}{\includegraphics{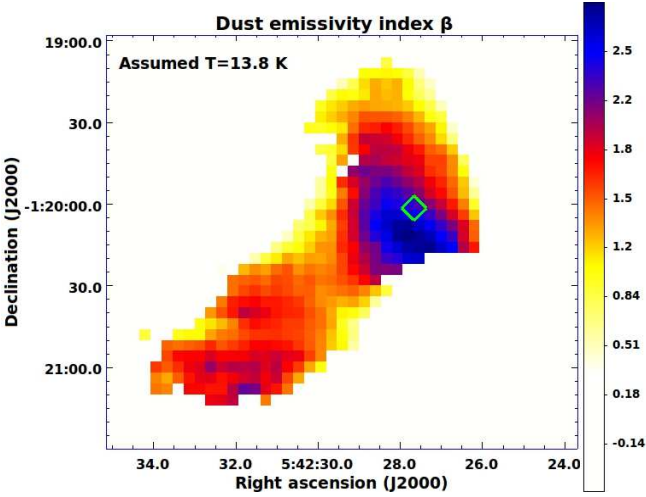}}
\caption{Distribution of the dust emissivity spectral index with the 
assumption that the dust temperature is constant at 13.8 K 
($=\langle T_{\rm kin} \rangle$; \cite{harju1993}). The diamond symbol 
indicates the 24-$\mu$m emission peak of I05399.}
\label{figure:beta}
\end{figure}

We can also estimate the value of the dust emissivity index by employing a 
constant value for the dust temperature (e.g., \cite{schnee2005}). 
As mentioned in Sect.~1., Harju et al. (1993) found that the average gas 
temperature in the I05399/SMM 1 -system is $13.8\pm3.6$ K, without significant 
variation between different parts within the errors (J.~Harju, private 
communication). Indeed, the $T_{\rm kin}$ values determined in Paper II towards 
I05399 and SMM 1 are both close to this average value. Therefore, it seems 
rather reasonable to assume that the average dust temperature of the system 
is about 13.8 K. The derived emissivity spectral index map is shown in 
Fig.~\ref{figure:beta}.

Measuring again the values towards the 350-$\mu$m peaks with aperture sizes 
corresponding to the core effective sizes, we obtain the average $\beta$ 
values of $2.2\pm0.4$ and $1.6\pm0.2$ for I05399 and SMM 1, respectively. 
These values are physically reasonable, and, within the errors, comparable to 
value $\beta=1.8$ adopted above.

\subsection{Core masses, and H$_2$ column and volume-averaged 
number densities}

The core masses, $M$, over an effective area of radius $R_{\rm eff}$
were estimated from the integrated 350- and 870-$\mu$m flux densities 
using the standard optically thin dust emission formulation [see,
e.g., Eq.~(2) in Paper I]. As the dust temperature 
we used the average values $22.3\pm10.2$ K and $12.5\pm1.2$ K for I05399 and 
SMM 1, respectively. Following the OH94 dust model described above (thin ice 
mantles), the dust opacity per unit dust mass at 350 and 870 $\mu$m was taken 
to be $\kappa_{{\rm 350\, \mu m}}=7.84$ cm$^2$~g$^{-1}$ and 
$\kappa_{{\rm 870\, \mu m}}=1.38$ cm$^2$~g$^{-1}$. The total dust-to-gas 
mass ratio was taken to be 1/141. The uncertainty in mass was 
propagated from the uncertainty in flux density, and for SMM 1 also from that 
of the temperature (the large std of $T_{\rm dust}$ for I05399 was not taken into 
account). The uncertainty in dust opacity, which is likely 
being a factor $\gtrsim2$, is the major source of error in the mass estimate.

Besides the values extracted from the $N({\rm H_2})$ map above, we also computed
the peak beam-averaged H$_2$ column densities from the peak surface 
brightnesses in a standard way [see, e.g., Eq.~(3) in Paper I]. 
The parameters needed in the calculation 
($T_{\rm dust}$, $\mu_{\rm H_2}$, $\kappa_{\lambda}$, $M_{\rm dust}/M_{\rm gas}$) 
were the same as described above. The error in $N({\rm H_2})$ is based 
on the uncertainty in the peak surface brightness, and for SMM 1 also on the 
$T_{\rm dust}$ uncertainty. Similarly to mass calculation, the $\kappa_{\lambda}$ 
uncertainty is likely to be the main source of error here.

The volume of a sphere-averaged H$_2$ number 
densities over $R_{\rm eff}$, $\langle n({\rm H_2}) \rangle$, were calculated 
using Eq.~(1) of Paper III, and the corresponding errors were propagated from 
those of $M$. The masses and densities derived above are listed in 
Cols.~(4)--(6) of Table~\ref{table:properties}. The core masses 
derived from 350-$\mu$m data are about half the values derived from the lower 
resolution 870-$\mu$m data. On the other hand, the column and number densities 
derived from 350-$\mu$m data are typically about three times higher compared 
to those derived from LABOCA data.

\begin{figure}[!h]
\centering
\resizebox{0.95\hsize}{!}{\includegraphics{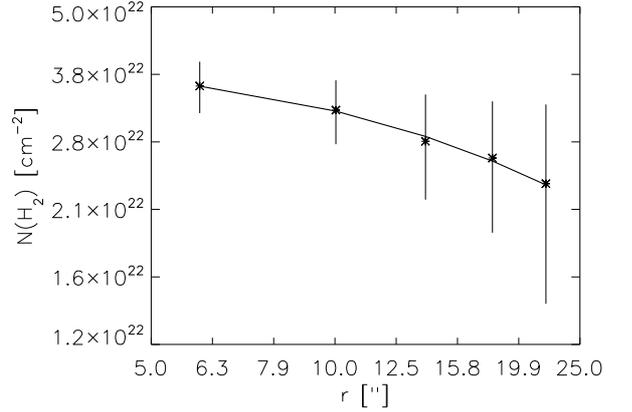}}
\caption{Radial profile of the column density of SMM 1 displayed with 
logarithmic scale for both axes. The data points represent azimuthally 
averaged values inside $4\arcsec$ wide annuli (Fig.~\ref{figure:dust}). 
The vertical error bars reflect the standard deviation of the azimuthal 
averaging. The mean radial $N({\rm H_2})$ profile can be fitted by a power-law 
with $N({\rm H_2})\propto r^{-\left(1.3_{-0.9}^{+2.2}\right)}$ between the flat 
inner core and the ``edge'', as indicated by the thick solid line.}
\label{figure:profile}
\end{figure}

We also constructed a column density distribution using azimuthally averaged 
values (see Fig.~\ref{figure:profile}). As shown in Fig.~\ref{figure:dust}, 
the $4\arcsec$- or one pixel-wide annuli were concentric 
circles centred on the $N({\rm H_2})$ peak of SMM 1 at $(\alpha,\, \delta)_{2000.0}=5^{\rm h}42^{\rm m}30\fs37,\,-1\degr 20\arcmin 32\farcs95$. The radial 
$N({\rm H_2})$ distribution is presented in Fig.~\ref{figure:profile}. 
In order to characterise the structure, the data points were 
fitted\footnote{We used the non-linear least squares IDL fitting routine MPFIT 
(\cite{markwardt2009}) available at {\tt http://purl.com/net/mpfit}} with 
the function [cf.~Eq.~(15) in \cite{nielbock2012}]

\begin{equation}
N({\rm H_2})=\frac{N({\rm H_2})_0}{1+\left(\frac{r}{r_0}\right)^m}\,,
\end{equation}
where $N({\rm H_2})_0$ is the central peak column density, $r$ is radial 
distance, $r_0$ is the radius of the flat inner region, and $m$ is the 
power-law exponent at large radii ($r \gg r_0$). The obtained values 
are $N({\rm H_2})_0=\left(4.1_{-1.1}^{+1.7}\right)\times10^{22}$ cm$^{-2}$, 
$r_0=28\farcs3_{-7\farcs6}^{+11\farcs6}\simeq0.06_{-0.01}^{+0.03}$ pc, and 
$m=1.3_{-0.9}^{+2.2}$. We note that $r_0$ is only roughly comparable 
to the thermal Jeans length at the centre of the 
annuli, i.e., $\lambda_{\rm J}\simeq 0.01\pm0.004$ pc at $T=11.9\pm0.9$ K 
(see, e.g., \cite{arzoumanian2011}; cf.~Sect.~4.5). In the case of a simple 
spherical source with a power-law volume (number) density profile of the form 
$n({\rm H_2})\propto r^{-p}$, the column density is expected to follow the 
relation $N({\rm H_2})\propto r^{-m}$ with $m=p-1$ (e.g., \cite{yun1991}). 
In our case, the density profile would then be 
$n({\rm H_2})\propto r^{-\left(2.3_{-0.9}^{+2.2}\right)}$.  
The deconvolved size of SMM 1, when determined from the LABOCA 
image [whose resolution equals that of the $N({\rm H_2})$ map], is very close 
to the beam size $19\farcs9$. Interestingly, when this fact is 
compared with the density power-law distribution models from Young et al. 
(2003; their Fig.~27), a value of $p \simeq 1.9$ would be expected, not far 
from the value deduced above (cf.~Sect.~5.3 in Paper I).

\begin{table*}
\caption{Dust colour temperatures, masses, beam-averaged peak H$_2$ co\-lumn 
densities, and volume-averaged H$_2$ number densities of the cores.}
\begin{minipage}{2\columnwidth}
\centering
\renewcommand{\footnoterule}{}
\label{table:properties}
\begin{tabular}{c c c c c c}
\hline\hline 
Source & $\langle T_{\rm dust}\rangle$ & $\lambda$ & $M$ & $N({\rm H_2})$ & $\langle n({\rm H_2}) \rangle$ \\  
       & [K] & [$\mu$m] & [M$_{\sun}$] & [$10^{22}$ cm$^{-2}$] & [$10^4$ cm$^{-3}$]\\ 
\hline
IRAS 05399-0121 & $22.3\pm10.2$ & 350 & $2.1\pm0.6$ & $8.9\pm2.7$ & $12.9\pm3.7$ \\                & & 870 & $3.8\pm0.4$ & $3.1\pm0.3$ & $3.8\pm0.4$ \\
SMM 1 & $12.5\pm1.2$ & 350 & $3.8\pm1.7$ & $7.0\pm3.2$ & $25.2\pm11.3$ \\
      & & 870 & $6.7\pm1.3$ & $3.8\pm0.8$ & $8.2\pm1.6$ \\
\hline 
\end{tabular} 
\end{minipage}
\end{table*}

\subsection{Modelling the spectral energy distribution of 
IRAS 05399-0121}

To characterise the physical properties of I05399, 
we constructed its SED by combining the \textit{Spitzer} photometric 
data with the submm data from SABOCA, LABOCA, and SCUBA. We also employed the 
\textit{IRAS} flux densities (see Table~1 in Paper I), for which a typical 
uncertainty of 10\% was adopted. Di Francesco et al. (2008) reported 
a rather large 850-$\mu$m effective radius for I05399, namely 
$49\farcs4$. The reported SCUBA flux densities are therefore likely to include 
contributions from SMM 1 also. For this reason, we used apertures of the 
size corresponding to the effective 350- and 870-$\mu$m core sizes to extract 
the 450- and 850-$\mu$m flux densities, respectively. The obtained values are 
$11.06\pm5.53$ and $2.59\pm0.52$ Jy, where the quoted uncertainties represent 
the 50\% and 20\% absolute flux uncertainties, respectively. 

To build the SED, we used the SED fitting tool developed 
by Robitaille et al. (2006, 2007)\footnote{The SED fitter is publicly 
available at {\tt http://caravan.astro.wisc.edu/protostars/}}. The tool 
employes a grid of two-dimensional axisymmetric radiative-transfer models of 
YSOs (\cite{whitney2003}) to fit the input flux densities. The grid 
encompasses a wide range of stellar masses (0.1--50 M$_{\sun}$; mainly 
covering low-mass YSOs) and YSO evolutionary stages (stellar ages of 
$10^3-10^7$ yr), and makes use of 20\,000 sets of 14 physical parameters and 
10 different viewing angles (inclinations ranging from edge-on to pole-on or 
from $90\degr$ to $0\degr$), resulting in a total of 200\,000 model SEDs. 
The models assume that the central embedded source is forming as a single 
star, it is associated with a flared disk, and that the system 
is surrounded by a dusty envelope with cavities carved by bipolar outflows. 
Here, a dust-to-gas ratio of 1/100 is assumed. 
The models take into account the fact that silicate dust can produce the 10- 
and 18-$\mu$m features (stretching of the Si-O bonds and bending of O-Si-O, 
respectively), but the models do not include the icy grain mantles or PAH 
emission features at mid-infrared. The model 
fitter requires the visual extinction and distance ranges within which to fit 
the observational data. We limited the former to be within the range 
$A_{\rm V}=0\ldots8.5$ mag, which is based on the median $A_{\rm V}$ value 
towards the NGC 2024 cluster (\cite{haisch2001}), and the  
distance was restricted to 450 pc .

\begin{figure}[!h]
\centering
\resizebox{0.95\hsize}{!}{\includegraphics{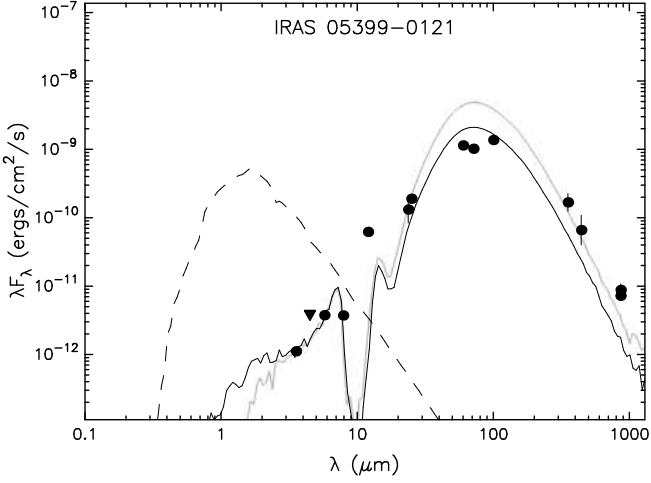}}
\caption{Spectral energy distribution (SED) of I05399. The filled black 
circles associated with vertical error bars represent the \textit{Spitzer}, 
\textit{IRAS}, SABOCA, SCUBA, and LABOCA flux densities.
The downwards pointing triangle indicates the \textit{Spitzer}/IRAC 4.5-$\mu$m 
flux density considered as an upper limit (not taken into account in the fit). 
The solid black line indicates the model that gives the 
best fit to the input flux densities, whereas the grey line 
illustrates the subsequent good fit model with the criterion 
$\chi^2-\chi_{\rm best}^2<5$, where $\chi^2$ is taken per data point 
($N_{\rm data}=13$). The black dashed curve shows the SED of the stellar 
photosphere corresponding to the central source of the best-fitting model 
(as it would appear in the absence of circumstellar dust, but including the 
effects of interstellar dust). Note the presence of silicate absorption 
features at 10 and 18 $\mu$m.}
\label{figure:SED}
\end{figure}

The SED is presented in Fig.~\ref{figure:SED}. 
%Its double-peaked appearance could be a sign of an edge-on disk (\cite{robitaille2007}). 
The best fit is shown by a black curve, whereas the grey curve shows 
the next-best fit satisfying the criterion 
$\chi^2-\chi_{\rm best}^2<5N_{\rm data}$, where $N_{\rm data}=13$ is the number of 
flux data points. The fit to the stellar 
photosphere SED is shown by a dashed line. Due to possible contamination by 
shock emission, the \textit{Spitzer}/IRAC 4.5-$\mu$m flux density was finally 
neglected from the fit (it was treated as an upper limit with a zero weight). 
Inclusion of this data point raised the luminosity to an unrealistically high 
level ($\sim230$ L$_{\sun}$). Some other IRAC bands could also be 
influenced by PAH emission or scattering: the 3.6, 5.8, and 8.0 $\mu$m 
bands contain PAH features at 3.3, 6.2, and 7.7 and 8.6 $\mu$m, respectively 
(\cite{draine2003}), and the 3.6-$\mu$m emission can originate from scattered 
interstellar radiation (\cite{steinacker2010}). The \textit{IRAS} 60- and 
100-$\mu$m flux densities appear to be slightly inconsistent with the 
\textit{Spitzer} 70-$\mu$m flux density measured by PSF fitting in Paper I. 
This is likely to result from the fact that the \textit{IRAS} beam of a few 
arcminutes picks up emission from the source surroundings.

The corresponding SED parameters are listed in Table~\ref{table:SED}. 
In this table, we give the inclination angle 
($i$), visual extinction between the observer and the outer edge of the 
circumstellar material ($A_{\rm V}$), stellar age, mass, and 
temperature ($\tau_{\star}$, $M_{\star}$, $T_{\star}$), the rate of mass 
accretion from the envelope onto the disk ($\dot{M}_{\rm env}$), disk mass 
($M_{\rm disk}$), visual extinction of the circumstellar material 
($A_{\rm V}^{\rm circ}$), total luminosity ($L_{\rm tot}$; contributions from 
the central star and accretion, assuming that all the accretion energy is 
radiated away), envelope mass ($M_{\rm env}$), and the implied evolutionary 
stage of the source. We give the best-fit model parameters with errors 
calculated as difference between the minimum and maximum values that can be 
derived from the three next-best model fits. We note that all the 
masses and the mass accretion rate assume a dust-to-gas ratio of 1/100.
The source inclination, $31\fdg8^{+17\fdg7}$, is roughly consistent with the 
observed outflow orientation. The best-fit model stellar mass and total 
luminosity are found to be $M_{\star}\sim0.5$ M$_{\sun}$ and $L_{\rm tot}\sim15.5$ 
L$_{\sun}$. The latter is corrected for foreground extinction and 
does not depend on the viewing angle. It is also comparable to the 
value $21\pm1.2$ L$_{\sun}$ derived by fitting a two-temperature component 
modified blackbody to the source SED (Paper I). The envelope mass, 
$M_{\rm env}\sim1.5^{+2.1}_{-0.9}$ M$_{\sun}$, determined from the long wavelength 
emission, is very similar to that derived from the 350-$\mu$m flux 
density alone, especially when the factor of 1.41 difference in the 
dust-to-gas ratio is taken into account. It is also comparable to the value 
$2.8\pm0.3$ M$_{\sun}$ derived for the cold SED component in Paper I (where 
the dust-to-gas ratio 1/100 was adopted and grains were assumed to 
have thick ice mantles).

In the case the range of values is very large, the parameter in question 
is not well constrained (such as the disk mass). I05399 can be classified 
following the YSO classification scheme of Robitaille et al. (2006). 
The evolutionary stage given in the last row of Table~\ref{table:SED} is 
based on the values of $\dot{M}_{\rm env}$ and $M_{\rm disk}$ relative to 
$M_{\star}$. I05399  has 
$\dot{M}_{\rm env}/M_{\star}\simeq 1.4\times10^{-4}>10^{-6}$ yr$^{-1}$, and is
therefore a Stage I candidate (dominated by a large accreting envelope). This 
stage is taken to encompass both the Stage 0 (i.e., 30-K modified black-body 
SED)\footnote{The Robitaille et al. (2006) SED models do not take into account 
the cold envelope material ($\lesssim30$ K), which might have some effect on 
the longer wavelength fits (see also \cite{offner2012b}). Therefore, the mass 
of the coldest envelope parts are neglected.} and I sources 
(\cite{robitaille2006}). The earliest evolutionary stage 0 is cha\-racterised 
by an envelope whose mass is higher than those of the central star and 
the disk (see \cite{mckee2010}). For I05399 we obtain the 
$M_{\rm env}/M_{\star}$ ratio of only $\sim3$, supporting the idea that it is at 
the Class (or Stage) 0/I borderline. Offner et al. (2012b) simulated 
the formation of low-mass stars including the effects of outflows and 
radiative heating. They found that while the SED models could correctly 
identify the evolutionary stage of the synthetic embedded protostars, the disk 
and stellar parameters inferred from the SEDs could be very different from the 
actual simulated values.
%In general, while SED-based source 
%classification can be dependable, for realistic protostellar sources with 
%outflows the mass and disk properties might be rather unreliable (\cite{offner2012b}). 
In particular, the fact that I05399 may be a binary 
system makes many of the derived source properties highly uncertain.

\begin{table}
\renewcommand{\footnoterule}{}
\caption{Results of SED modelling of IRAS 05399-0121.}
{\tiny
\begin{minipage}{1\columnwidth}
\centering
\label{table:SED}
\begin{tabular}{c c}
\hline\hline 
Parameter & Value\tablefootmark{a}\\
\hline
Min. $\chi^2/N_{\rm data}$ & $103.6^{+36.3}$ \\[1ex]
Inclination, $i$ [$\degr$] & $31.8^{+17.7}$ \\[1ex] 
Visual extinction, $A_{\rm V}$ [mag] & $2.2^{+2.2}_{-1.4}$ \\[1ex]
Stellar age, $\tau_{\star}$ [$10^3$ yr] & $6.3^{+8.4}_{-5.2}$ \\[1ex]
Stellar mass, $M_{\star}$ [M$_{\sun}$] & $0.46^{+1.09}$ \\[1ex]
Stellar temperature, $T_{\star}$ [$10^3$ K] & $3.65^{+0.53}$\\[1ex]
Envelope accretion rate, $\dot{M}_{\rm env}$ [$10^{-5}$ M$_{\sun}$~yr$^{-1}$] & $6.47^{+9.13}_{-4.35}$ \\[1ex]
Disk mass, $M_{\rm disk}$ [M$_{\sun}$] & $0.0352^{+0.1278}_{-0.0345}$ \\[1ex]
Circumstellar extinction, $A_{\rm V}^{\rm circ}$ [mag] & $430^{+376}_{-29}$ \\[1ex]
Total luminosity, $L_{\rm tot}$ [L$_{\sun}$] & $15.5^{+23.6}_{-8.62}$ \\[1ex]
Envelope mass, $M_{\rm env}$ [M$_{\sun}$] & $1.51^{+2.07}_{-0.91}$ \\[1ex]
Stage & I \\[1ex]
\hline 
\end{tabular} 
\tablefoot{\tablefoottext{a}{For each model parameter, the best-fit value is 
given with the $\pm$ errors computed from the minimum and maximum 
values obtained from the three next-best model fits. For some parameters, the 
best-fit value is the lowest one.} }
\end{minipage} }
\end{table}

%\begin{table}
%\renewcommand{\footnoterule}{}
%\caption{Results of SED modelling of IRAS 05399-0121.}
%{\tiny
%\begin{minipage}{1\columnwidth}
%\centering
%\label{table:SED}
%\begin{tabular}{c c}
%\hline\hline 
%Parameter & Value\tablefootmark{a}\\
%\hline
%Min. $\chi^2/N_{\rm data}$ & 103.6 [108.3--139.9] \\
%Inclination, $i$ [$\degr$] & 31.8 [41.4--49.5] \\ 
%Visual extinction, $A_{\rm V}$ [mag] & 2.2 [0.8--4.4] \\
%Stellar age, $\tau_{\star}$ [$10^3$ yr] & 6.3 [1.1--14.7] \\
%Stellar mass, $M_{\star}$ [M$_{\sun}$] & 0.46 [0.48--1.55] \\
%Stellar temperature, $T_{\star}$ [$10^3$ K] & 3.65 [3.66--4.18]\\
%Envelope accretion rate, $\dot{M}_{\rm env}$ [$10^{-5}$ M$_{\sun}$~yr$^{-1}$] & 6.47 [2.12--15.6] \\
%Disk mass, $M_{\rm disk}$ [M$_{\sun}$] & 0.0352 [$7.4\times10^{-4}$--0.163]\\
%Circumstellar extinction, $A_{\rm V}^{\rm circ}$ [mag] & 430 [401--806] \\
%Total luminosity, $L_{\rm tot}$ [L$_{\sun}$] & 15.5 [6.88--39.1] \\
%Envelope mass, $M_{\rm env}$ [M$_{\sun}$] & 1.51 [0.60--3.58] \\
%Stage & I \\
%\hline 
%\end{tabular} 
%\tablefoot{\tablefoottext{a}{For each model parameter, the best-fit value is 
%given, followed by a range defined by the minimum and maximum values obtained 
%from the three next-best model fits.} }
%\end{minipage} }
%\end{table}

\subsection{Fragmentation of the source}

To investigate the dynamical state of the detected 350-$\mu$m filamentary 
structure as a whole, we estimated its total mass as

\begin{equation}
M=\mu_{\rm H_2}m_{\rm H}\sum_i \left[N({\rm H_2})A_{\rm pix}\right]_i\,,
\end{equation}
where $m_{\rm H}$ is the mass of a hydrogen atom, $A_{\rm pix}$ is the surface 
area in one pixel ($16 \,\square\arcsec$), and the sum is over all pixels 
within the $3\sigma$ contour of 350-$\mu$m emission. This integration of the 
mass surface density gives $M\simeq 8.8$ M$_{\sun}$. As the projected 
length of the source along the long axis is 0.26 pc, the corresponding mass 
per unit length, or line mass, is $M_{\rm line}\simeq 34$ 
M$_{\sun}$~pc$^{-1}$.

As having a filamentary shape, the I05399/SMM 1 -system can be considered a 
cylindrically shaped object. It is therefore interesting to examine if the 
source's substructure can be understood in the context of cylindrical 
fragmentation. For an infinite, unmagnetised, isothermal 
cylinder, the instability (collapse to line singularity) is reached if its 
$M_{\rm line}$ exceeds the critical equilibrium value of (e.g., 
\cite{ostriker1964}; \cite{inutsuka1992})

\begin{equation}
\label{eq:linemass}
M_{\rm line}^{\rm crit}=\frac{2c_{\rm s}^2}{G}\,,
\end{equation}
where $G$ is the gravitational constant. Using 
the average gas temperature of $\langle T_{\rm kin}\rangle=13.8\pm3.6$ K 
(\cite{harju1993}) and the mean molecular weight per free particle of 
$\mu_{\rm p}=2.37$ (\cite{kauffmann2008})\footnote{The classical value 
of $\mu_{\rm p}=2.33$ applies for a gas consisting of H$_2$ molecules with 
10\% He (He/H$=0.1$) and a negligible amount of metals.} to calculate the 
sound speed, we derive the value 
$M_{\rm line}^{\rm crit}\simeq 22\pm6$ M$_{\sun}$~pc$^{-1}$. 
The I05399/SMM 1 -system therefore appears to be a thermally 
supercritical filament by a factor of $1.5\pm0.4$. If the external pressure 
is not negligible, as may well be the case for the dynamic Orion B9 region, 
then $M_{\rm line}^{\rm crit}$ is smaller than the above value (e.g., 
\cite{fiege2000}). On the other hand, if the non-thermal motions are taken 
into account, i.e., $c_{\rm s}$ is replaced by the effective sound speed 
$c_{\rm eff}=(c_{\rm s}^2+\sigma_{\rm NT}^2)^{1/2}$, where 
$\sigma_{\rm NT}\sim c_{\rm s}$ (Paper II), the value of $M_{\rm line}^{\rm crit}$ 
becomes about two times larger than in Eq.~(\ref{eq:linemass}) 
(\cite{fiege2000}). We would like to stress that this includes the assumption 
that all of the non-thermal motions are providing additional support, 
similarly to that provided by thermal pressure (\cite{kirk2013}). 
The fact that our source already contains a protostellar object, i.e., is 
collapsing, suggests that it should be (slightly) supercritical. 
We note that the source mass determined from the $N({\rm H_2})$ map suffers 
from uncertainty in the dust properties (such as the dust-to-gas ratio and 
dust opacity).

As mentioned in Sect.~3.1, the projected separation between the subcores 
I05399 and SMM 1 is 0.14 pc at $d=450$ pc (the spacing ranges from 0.13 to 
0.16 pc for the distance range 400--500 pc). In the case the cylinder has 
$M_{\rm line}=M_{\rm line}^{\rm crit}$ as is roughly the case for our source, 
the Jeans length along the long axis is (\cite{larson1985}; \cite{hartmann2002})

\begin{equation}
\lambda_{\rm J}^{\rm c}=\frac{3.94c_{\rm s}^2}{G\Sigma_0}\,,
\end{equation}
where $\Sigma_0=\mu_{\rm H_2}m_{\rm H}N({\rm H_2})$ is the central surface 
density. We estimated the central column density to be the 
average value along the slice shown in Fig.~\ref{figure:dust} 
(position angle $135\fdg6$ east of north), i.e., 
$\sim 2.6\pm1.0\times10^{22}$ cm$^{-2}$, where the uncertainty is 
given by one standard deviation. The corresponding value of 
$\lambda_{\rm J}^{\rm c}$ is $0.075\pm0.035$ pc. This is about 
$1.9\pm0.9$ times smaller than the observed value but within the errors they 
are in reasonable agreement. The system may therefore lie close to the plane 
of the sky. For comparison, the traditional thermal Jeans length is

\begin{equation}
\lambda_{\rm J}=\sqrt{\frac{\pi c_{\rm s}^2}{G\langle \rho \rangle}}\,,
\end{equation}
where $\langle \rho \rangle=\mu_{\rm H_2}m_{\rm H}\langle n({\rm H_2})\rangle$ is 
the mean mass density (e.g, \cite{kauffmann2010}). If we use the average 
density of $\sim 6\pm1\times10^4$ cm$^{-3}$ derived from the 870-$\mu$m values 
listed in Col.~(6) of Table~\ref{table:properties}, we obtain 
$\lambda_{\rm J}\simeq 0.09\pm0.01$ pc. This is also roughly 
comparable to the observed core separation.

If the \textit{finite} length and diameter of the cylindrical cloud are $L$ 
and $D$, respectively, the number of fragments forming due to gravitational 
instability is expected to be (e.g., \cite{bastien1991}; \cite{wiseman1998})

\begin{equation}
N_{\rm frag}=\frac{L}{\lambda_{\rm crit}}=\frac{2\left(L/D \right)}{3.94}\simeq0.5A\,,
\end{equation}
where $\lambda_{\rm crit}=1.97D$ is the wavelength of the most unstable 
perturbation, and $A$ the cylinder's aspect ratio. As mentioned earlier, $A$ 
is about 4, so one would expect to have two subfragments, in agreement with 
the two subcores I05399 and SMM 1. Note that the average width of the SABOCA 
filament, $D\sim0.07$ pc, yields $\lambda_{\rm crit}\simeq0.14$ pc, 
which is in excellent agreement with the observed separation distance.

To conclude, the studied system has likely fragmented into subcores as a 
result of cylindrical gravitational instability. Considering a filament of 
radius $R$, the fragmentation timescale is then expected to be comparable to 
the radial crossing time, $\tau_{\rm cross}=R/\sigma$, where $\sigma$ refers to 
the total velocity dispersion. If we use the average FWHM linewidth of 
$\Delta {\rm v}=\sqrt{8 \ln 2}\sigma=0.6$ km~s$^{-1}$ from Harju et al. (1993), 
which is very similar to the NH$_3(1,\,1)$ linewidths found in Paper II, the 
$\tau_{\rm cross}$ for the SABOCA filament is $\sim1.34\times10^5$ yr. 
If I05399 has accreted mass at a constant rate since the formation of 
the central protostar, the fragmentation timescale is $21_{-12}^{+101}$ times 
longer than the protostar's age inferred from SED modelling 
(Table~\ref{table:SED}).

%As having a filamentary shape, the I05399/SMM 1 -system can be considered a 
%cylindrically shaped object. For an infinite cylindrical isothermal cloud, 
%gravitational fragmentation will produce a chain of condensations along the 
%cylinder, whose periodic separation corresponds to the wavelength of the most 
%unstable perturbation (e.g., \cite{tomisaka1995})

%\begin{equation}
%\lambda_{\rm max}\simeq \frac{20 c_{\rm s}}{\sqrt{4 \pi G \rho_{\rm c}}}\simeq \frac{5.64 c_{\rm s}}{\sqrt{G \rho_{\rm c}}}\,,
%\end{equation}
%where $\rho_{\rm c}$ is the central density. The value of the latter quantity 
%is unclear but $\sim10^5$ cm$^{-3}$ seems reasonable given the 
%average densities of the 350-$\mu$m cores. This translates into the value 
%$\lambda_{\rm max}\simeq0.25$ pc, which exceeds the observed projected 
%separation by a factor of 1.8.
%for a random distribution of inclination angles, $i\approx 57\degr$, by a factor of $1/\cos(i)\simeq1.84$; correction factor
%would match the length if the inclination was

\section{Discussion}

\subsection{Dust properties}

As explained in Sect.~4.1, the temperatures measured for the gas component 
in Paper II are similar to the dust temperatures measured here from the 
350-to-870-$\mu$m flux density ratios. The similarity between $T_{\rm kin}$ and 
$T_{\rm dust}$ suggests that the gas and dust are well-coupled. This is not 
necessarily surprising, because thermal coupling between the gas and dust by 
collisions is expected at densities $n \gtrsim10^4$ cm$^{-3}$ 
(\cite{goldsmith1978}).

The constructed $T_{\rm dust}$ map shown in Fig.~\ref{figure:dust} reveals a 
warm spot towards I05399. Given the presence of an embedded protostar, the 
increased 350/870-$\mu$m flux density ratio is likely dominated by 
temperature, although it could be partly due to variation in $\beta$ also.
However, the 24-$\mu$m peak position and the warmest 
spot are not exactly coincident, but there is a slight offset of about 
$9\arcsec$ between the two. This difference could be caused by the outflow 
feedback from I05399, namely heating by outflow shocks. 
Also, the $T_{\rm dust}$ values appear to rise when going southeast of the 
350-$\mu$m peak of SMM 1. Perhaps this is another reflection of the outflow 
interaction related to the HH 92 jet. This conforms to the low CO depletion 
factors and to the detection of deuterated formaldehyde (D$_2$CO; a species 
forming on icy grain surfaces) reported in Paper III as described in Sect.~1. 
It is worth noting, however, that the warmer southern parts of the 
I05399/SMM 1 -system face the direction towards the Flame Nebula NGC 2024, 
an active massive star-forming region (see Fig.~9 in \cite{miettinen2012}). 
Even though this could enhance the strength of the local interstellar 
radiation field and thus provide external dust heating, one would then expect 
to see a more uniform $T_{\rm dust}$ gradient across the short axis and along 
the whole source. Therefore, protostellar or shock heating seems more plausible 
here.

By fixing the value of $T_{\rm dust}$ to 13.8 K, i.e., to the average 
$T_{\rm kin}$ derived by Harju et al. (1993), we also computed the map of dust 
emissivity spectral index, $\beta$ (Fig.~\ref{figure:beta}). The $\beta$ 
varies from $\sim1.3$ to 2.8 inside the $3\sigma$ SABOCA contour. For 
comparison, through a multiwavelength study of the starless core TMC-1C, 
Schnee et al. (2010) derived the values $1.7\leq \beta \leq 2.7$, the most 
likely one being $\sim2.2$. Shirley et al. (2011), by studying the Class 0 
source B335, found a relatively steep submm opacity power-law index of 
$\beta=(2.18-2.58)^{+0.30}_{-0.30}$. From Fig.~\ref{figure:beta} we see that 
$\beta$ appears to rise to values $>2$ towards those parts where 
$T_{\rm dust}$ is increased. Of course, our present data does not allow to draw 
any firm conclusions and one must be careful not to overinterpret the current 
results, but it can still be speculated that the increase of 
$\beta$ is caused by the destruction of dust grains. This would be in line 
with the destructive outflow feedback discussed above. Depending on 
the velocity of the jet-induced shock, dust grains can be either partially or 
completely destroyed by sputtering, grain-grain collisions, and shattering 
(e.g., \cite{jones1994}; \cite{vanloo2013}). Moreover, the processing by both 
UV photons and cosmic rays can cause non-thermal desorption of volatile 
species from the icy grain mantles into the gas phase 
(e.g., \cite{hartquist1990}; \cite{roberts2007}). 
However, it is also possible that higher values of $\beta$ are actually caused 
by the growth of icy mantles on dust grains (e.g., \cite{schnee2010a} and 
refe\-rences therein), but this seems to contradict the chemical 
features discussed earlier. Conversely, observational results of 
small values of $\beta$ are often interpreted to be indicative of grain growth 
(e.g., \cite{testi2003}; \cite{draine2006}; \cite{kwon2009}). 
For example, Kwon et al. (2009) found that for their sample of Class 0 sources 
$\beta\lesssim1$, implying larger grain sizes due to 
gas accretion and/or coagulation. We note, however, that a visual inspection 
of the \textit{Spitzer}/IRAC 3.6-$\mu$m image of I05399 reveals the presence 
of a coreshine-like emision (Fig.~\ref{figure:jet}). The 3.6-$\mu$m coreshine 
originates in dust-grain scattering of the background radiation, and is an 
indication of the presence of large, micron-size grains whose  
abundance is roughly given by the MRN (\cite{mathis1977}) grain-size 
distribution (\cite{steinacker2010}). For example, Stutz et al. 
(2010) found this towards the Bok globule CB 244 (coincidentally, it resembles 
our source in the sense that it contains a Class 0 protostar and a starless 
core). Finally, the anti-correlation between $T_{\rm dust}$ 
and $\beta$ claimed in some observational studies might be artificial and 
caused by noise effects and $T_{\rm dust}$--$\beta$ degeneracy instead of being 
a real physical trend (e.g., \cite{kelly2012} and references therein).

\subsection{IRAS 05399-0121/SMM 1 - a double core system}

One of the original aims of the present high-resolution SABOCA study was to 
search for substructures within I05399 and SMM 1. The fact that SMM 1 remains 
a single source is consistent with the finding by Schnee et al. (2010b) that 
none of their 11 starless cores break into smaller components at $5\arcsec$ 
resolution. Although, higher interferometric resolution may be 
necessary to see substructure at such scales (\cite{offner2012a}).

The I05399/SMM 1 -system appears to be a filamentary-type object not far from 
equilibrium (slightly thermally supercritical), and the origin of the 
two subcores, projectively separated by 0.14 pc, seems to be caused by 
cylindrical Jeans-type fragmentation. We have not applied a 
correction factor for inclination, but the long length of the outflow driven 
by I05399, and its possible interaction with SMM 1 suggest that the system is 
aligned close to the plane of the sky.

Although the fragmentation of the studied system can be explained in the 
context of a self-gravitating isothermal equilibrium filament, the estimated 
density profile $n(r)\propto r^{-\left(2.3_{-0.9}^{+2.2}\right)}$, 
albeit quite uncertain, is shallower than what would be theoretically 
expected, i.e., $\sim r^{-4}$ (\cite{ostriker1964}; see also 
\cite{arzoumanian2011}; \cite{pineda2011}). Instead, it is close to the 
standard singular isothermal sphere profile of $n(r)\propto r^{-2}$ 
(\cite{shu1977}). However, Fiege \& Pudritz (2000) 
found that most of their models of isothermal filaments that are 
threaded by helical magnetic fields and are in gravitational virial 
ba\-lance, can yield density profiles that decline like $\sim r^{-1.8}$ 
to $\sim r^{-2}$ (see also \cite{lada1999}). Similarly, 
Caselli et al. (2002) found that starless cores can be modelled with radial 
density profiles of $n(r)\propto r^{-2}$ outside $\sim0.03$ pc, whereas 
protostellar cores are better described with steeper single power-law density 
profiles ($p \geq2$). For comparison, Shirley et al. (2002) modelled seven 
Class 0 objects using a single power-law density profiles and found that the 
best-fit power-law index is $p=1.8\pm0.1$. The same density profile ($p=1.8$) 
was found by Schnee \& Goodman (2005) for TMC-1C through SCUBA 450/850 $\mu$m 
observations, resembling our 350/870 $\mu$m analysis.

If both I05399 and SMM 1 were formed simultaneously through fragmentation,  
it is interesting why they are at so different stages of evolution: the 
former is already a relatively evolved protostar, while the latter one is 
still starless. This is not a unique case in this sense; a 
similar situation holds for the double-core globules CB 26 and 244 recently 
studied by Launhardt et al. (2013). Also, the filamentary clump 
associated with the Class 0 protostar IRAS 05405-0117 in Orion B9 shows 
starless substructure in the N$_2$H$^+(1-0)$ map by Caselli \& Myers (1994) 
and in submm dust emission maps (Papers I and III). 
Miettinen (2012) suggested that the 
core formation in Orion B9 could have been triggered by the feedback from 
NGC 2024. Externally induced compression could have led to a phase of rapid 
accretion (\cite{hennebelle2003}; \cite{motoyama2003}), 
which might be able to explain the advanced evolutionary stage of I05399, 
but it is unclear why SMM 1 would have not experienced the same fate. 
The length of the outflow driven by I05399 supports the 
hypothesis that the central source is a strong accretor. Because I05399 lies 
at the tip of both the major parent filament (see Fig.~9 in 
\cite{miettinen2012}) and the smaller scale elongated structure studied here, 
it could represent a point where gravitational focusing has enabled the rapid 
evolution into protostellar phase (\cite{tobin2010}). Again, why this would 
have not happened for SMM 1 remains unclear.

The virial mass we derived for SMM 1 in Paper II is $M_{\rm vir}=11.1\pm0.3$ 
M$_{\sun}$, where it was assumed that $n(r)\propto r^{-1}$. If this is compared 
with the revised 870-$\mu$m core mass derived here, we obtain the virial 
parameter $\alpha_{\rm vir}=M_{\rm vir}/M=1.7\pm0.3$ (\cite{bertoldi1992}). 
If the density profile of SMM 1 is of the form $n(r)\propto r^{-2.3}$, 
the above $M_{\rm vir}$ should be divided by a 
factor of 2.62, and, as a cigar-shaped prolate spheroid with the axial 
ratio of about 1.8, a further division by a factor of $\sim0.8$ 
should also be applied (\cite{bertoldi1992}; \cite{schnee2005}). These 
correction factors yield the value $\alpha_{\rm vir}=0.8\pm0.2$. 
SMM 1 is therefore very likely to be gravitationally bound 
($\alpha_{\rm vir}<2$) and near virial equilibrium ($\alpha_{\rm vir}=1$).
Also the high density and degree of deuterium fractionation in SMM 1 support 
its prestellar nature. The fact that I05399 has collapsed faster than SMM 1 
could mean that the latter has stronger internal forces resisting gravity. 
One might also conclude that the outflow from I05399 has triggered the 
formation of SMM 1, but that it has not collapsed to form a protostar yet.

%Models of core formation leading to inside-out collapse
%predict a Ñat inner core approaching $p\sim2$ at larger radii

%before the core has fully
%relaxed to a singular isothermal sphere with n(r) r-2 (e.g.,
%Shu 1977).

%The inside-out collapse model of a singular isothermal sphere
%(Shu 1977) predicts a broken power-law density distribution
%with an $r-2$ profile outside the infall radius and a profile inside
%the infall radius asymptotically approaching $r-1.5$.; roughlöy agreement with some AD models

%The theoretical model of inside-out collapse of prestellar
%cores predicts a density profile of nH  r-2 (e.g. Shu 1977; Shu
%et al. 1987),

%see eq 14 in harjus new paper to predict the value of $r_0$

\section{Summary and conclusions}
   
We have mapped the dense core system IRAS 05399/SMM 1 in Orion B9 at 350 and 
870 $\mu$m using the bolometer cameras SABOCA and LABOCA on APEX. The present 
study demonstrates the complementarity of these two instruments, similarly to 
what has been done in other studies using SCUBA at 450 and 850 $\mu$m on JCMT.
In particular, the present data allowed us to construct the maps of dust 
colour temperature and column density at relatively high angular resolution 
($19\farcs9$).

The source is filamentary in shape, and there is a protostellar core at 
the nortwestern tip (I05399) and a starless core at the southeastern end 
(SMM 1). By comparing the total line mass of the SABOCA filament to its 
critical equilibrium value, we conclude that as a whole it is slightly 
thermally supercritical. The observational fact that there appears to be two 
subcores projectively separated by 0.14 pc can be understood in terms of 
Jeans-type cylindrical fragmentation. A simple analysis suggests that the 
radial density profile is shallower than for an isothermal unmagnetised 
filament in equilibrium ($\sim r^{-\left(2.3_{-0.9}^{+2.2}\right)}$ vs. 
$\sim r^{-4}$).

The SED we constructed for I05399 suggests that it 
is near the transition phase between Stages 0 and I, in agreement with our 
previous results. To our knowledge, we have presented the first 
\textit{Spitzer}/IRAC images of the spectacular outflow driven by I05399. The 
quadrupolar-type jet morphology seen in the 3.6-$\mu$m image supports the 
possibility that I05399 is actually a binary protostar.

If I05399 and SMM 1 were formed simultaneously through fragmentation of the 
parent filament, the distinct evolutionary stage between the two could be the 
result of rapid mass accretion onto I05399 in a high-pressure environment. 
In principle, this could be the result of core formation triggered 
by external feedback from NGC 2024 as suggested by Miettinen (2012), 
but why SMM 1 would not have experienced the same effect is unclear.
The temperature map we computed (under the assumption of constant dust opacity 
spectral index $\beta=1.8$) shows a few warmer parts, one associated with 
I05399 and the other at the southeastern end. These could, in part, be 
imprints of the outflow feedback from I05399, as supported by our earlier 
results on the chemical properties of the source (e.g., low CO depletion). 
However, it cannot be excluded that NGC 2024, some $\sim30\arcmin$ 
($\sim4$ pc) southwest of I05399/SMM 1, is proving external heating, although 
the source's temperature gradient does not support this idea. 

One of our original aims here was to search for substructures or dense 
condensations within the cores. Given the results presented in Paper III 
(many of the Orion B9 cores are split into substructures consistent with Jeans 
instability), the high spatial resolution of the present SABOCA data, 
$\sim3\,400$ AU, could have revealed some further sub-fragmentation. 
However, the cores I05399 and SMM 1 do not break up into sub-condensations but 
the system remains a double source at the current resolution.

\begin{acknowledgements}

We thank the referee, Paola Caselli, for a thorough reading of the 
manuscript and providing helpful comments and suggestions.
We are grateful to the staff at the APEX telescope for 
performing the service-mode SABOCA/LABOCA observations presented in this 
paper. O.~M. acknowledges the Academy of Finland for the financial support 
through grant 132291. S.~S.~R.~O. acknowledges support from NSF Fellowship 
AST-0901055 and NASA through Hubble Fellowship grant HF-51311. 
This work is based in part on observations made with the 
\textit{Spitzer Space Telescope}, which is operated by the Jet Propulsion 
Laboratory, California Institute of Technology under a contract with NASA. 
This research has made use of NASA's Astrophysics Data System and the 
NASA/IPAC Infrared Science Archive, which is operated by the JPL, California 
Institute of Technology, under contract with the NASA.
  
\end{acknowledgements}

\end{document}